\documentclass[11pt,a4paper]{article}
\pdfoutput = 1
\usepackage{jcappub}
\usepackage[utf8]{inputenc}
\usepackage{amsmath,amssymb}
\usepackage{graphicx}
\usepackage{hyperref}
\usepackage{natbib}

\subheader{\footnotesize
        \vspace*{-3.7em}
        \begin{flushright}
                CERN-TH-2020-115
        \end{flushright}
}

\newcommand{\edit}[1]{{{{#1}}}}

\newcommand{\bp}{\textbf{p}}
\newcommand{\bq}{\textbf{q}}

\newcommand{\bk}{\textbf{k}}

\newcommand{\bPsi}{\boldsymbol{\Psi}}

\def \dtq{\int d^3 \bq \ }

\author[a]{Shi-Fan Chen}
\author[b]{Zvonimir Vlah}
\author[a]{Martin White}

\affiliation[a]{Department of Physics, University of California,
Berkeley, CA 94720}
\affiliation[b]{Theory Department, CERN, CH-1211 Geneve 23, Switzerland}

\emailAdd{shifan\_chen@berkeley.edu}
\emailAdd{zvonimir.vlah@cern.ch}
\emailAdd{mwhite@berkeley.edu}

\title{Modeling features in the redshift-space halo power spectrum with perturbation theory}


\abstract{We study the ability of perturbative models with effective field theory contributions and infra-red resummation to model the redshift space clustering of biased tracers in models where the linear power spectrum has ``features'' -- either imprinted during inflation or induced by non-standard expansion histories.  We show that both Eulerian and Lagrangian perturbation theory are capable of reproducing the Fourier space two-point functions of halos up to the non-linear scale from a suite of $4096^3$ particle N-body simulations.  This is the first demonstration that perturbative models can accurately fit the redshift-space clustering of biased tracers in N-body simulations of such theories.  By comparing different theoretical models and IR resummation schemes we assess the current theoretical uncertainty in predicting power spectra for models with features. Our results suggest that future surveys will be able to detect or tightly constrain features in the primordial spectrum below the one percent level across a wide range of scales.}
\begin{document}
\maketitle
\flushbottom

\section{Introduction}

Current observations of large-scale structure are consistent with a primordial power spectrum that is featureless, upon which $14\,$Gyr of evolution imprints two characteristic scales: the horizon at the epoch of matter-radiation equality and the sound horizon of photon-baryon acoustic oscillations prior to recombination \cite{Peacock99,Dodelson03,PlanckLegacy18}.  However many modifications of the standard model would lead to deviations from this simple picture.  Many inflationary models imprint features in the otherwise smooth spectrum of curvature fluctuations at early times (see e.g.\  refs.~\cite{Chluba15,Slosar19b} for recent reviews and references to the extensive early literature).  While these features are in some sense generic, different models predict very different properties for these features, their bandwidth, amplitude and shape.  Detection of such features would open new windows into the primordial Universe.
In addition, the evolution of these primordial perturbations across $14\,$Gyr of cosmic history can `induce' features in the observed spectrum.  One class of features which has been the focus of intense observational activity are baryon acoustic oscillations \cite{Weinberg13}.  However more generally new types of particle interactions, new energy components or changes in the expansion history can all alter the observed, late-time spectrum in observable ways.

The strongest constraints on primordial features to date come from a combination of cosmic microwave background anisotropies \cite{PlanckLegacy18,PlanckInf18} and large-scale structure \cite{Beutler19}.  This has placed upper limits on the amplitude of features at the several percent level for frequencies in the range $10^2<\omega<10^3\,h^{-1}{\rm Mpc}$ (see previous references for more details).  Future large-scale structure surveys, especially those performed at high redshift over large cosmic volumes, should be able to tighten these constraints significantly \cite{Slosar19b}.

In this paper we investigate the degree to which modern perturbative calculations can quantitatively predict the clustering of biased tracers in redshift space, including the ``washing out'' of primordial features by mode-coupling associated with non-linear evolution, changes to the broad-band shape of the spectrum by non-linear biasing and the mixing of the velocity and density perturbations through redshift-space distortions.  We are not the first authors to address these topics, indeed there is an extensive literature on this topic within the context of baryon acoustic oscillation (BAO) studies (see \S\ref{sec:PT} for references).  An important finding of these studies is the existence of $\mathcal{O}(1)$ corrections to features for a wide range of parameters and wavenumbers of interest.  This leads one to consider resumming these $\mathcal{O}(1)$ contributions, which arise from the large displacements, a process referred to as IR resummation.  One method of deriving the IR resummation process is as a saddle-point approximation to a particular integral (\S\ref{sec:PT}).  In this paper we pay particular attention to the manner in which IR resummation, mode coupling and the mixing of density and velocity perturbations implied by redshift-space distortions appear in models with features at different scales than BAO, and how IR resummation can deal with features with $k$-dependent frequency where the choice of saddle is not immediately obvious.

The outline of the paper is as follows.  In Section \ref{sec:nbody} we introduce the particle-mesh simulations that we use to validate our perturbative models.  Section \ref{sec:models} describes the specific feature models that we test, which have been chosen to be representative of different classes that appear in the literature, while Section \ref{sec:PT} describes the perturbative calculations we investigate.  Section \ref{sec:results} presents the comparison between the theory and N-body power spectra and we conclude in Section \ref{sec:conclusions}.  Throughout we will assume a $\Lambda$CDM cosmological model consistent with the latest {\sl Planck} results \cite{PlanckLegacy18,PCP18} and quote distances in comoving $h^{-1}$Mpc.

\section{N-body simulations}
\label{sec:nbody}

To validate our model and further investigate the impact of non-linearity, bias and redshift-space distortions on primordial features we have run a number of N-body simulations.  For each of several models we generated 6 realizations of Gaussian initial conditions at $z=9$ using $2^{\rm nd}$ order Lagrangian perturbation theory and employed the {\sc Fastpm} code \cite{Feng19} to evolve $4096^3$ particles in a $2.5\,h^{-1}$Gpc box with a $(3\times 4096)^3$ force grid over 40 time steps linearly spaced in the scale factor, $a$, down to $z=0.5$.  With 40 steps, which improves the convergence at higher $k$, the code behaves very much as a traditional particle mesh code.

Each model employed the same background cosmology, of the $\Lambda$CDM family and consistent with the latest constraints from {\sc Planck} \cite{PCP18}.  Halo catalogs and 5 per cent of the dark matter particles were output at $z=2$, 1 and $0.5$.  The density power spectra in real and redshift space and the real-space velocity spectra were computed using the {\sc Nbodykit} software \cite{nbdkit}.  Fourier transforms were done on a $4096^3$ mesh.  We bin the spectra in linear $k$ bins of width $\Delta k=0.005\,h\,{\rm Mpc}^{-1}$ starting at $k_{\rm min}=0.005\,h\,{\rm Mpc}^{-1}$.  We compute power spectrum ``wedges'', $P(k,\mu)$, in 5 equal width $\mu$ bins centered at $\mu=0.1$, 0.3, 0.5, 0.7 and 0.9, as well as power spectrum multipoles, $P_\ell(k)$ for $\ell=0$, 2 and 4.  \edit{We use the plane-parallel approximation throughout this work, for a periodic box within which estimating $P(k,\mu)$ amounts to simply Fourier-transforming a gridded field, squaring, and binning by $k$ and $\mu$ without the usual observational complications involving window functions or line-of-sight effects which would need to be accounted for in real surveys like BOSS \cite{Grieb17}.} We do not remove shot noise from any of our spectra, but rather include such contributions in our models.  We present the average of the $P(k,\mu)$ with the line of sight taken along each of the cardinal directions of the box.

We have chosen to focus on two, mass limited halo samples, with densities of $10^{-3}\,h^3\,{\rm Mpc}^{-3}$ and $10^{-4}\,h^3\,{\rm Mpc}^{-3}$. These are characteristic of densities achieved by surveys such as DESI \cite{DESI}, MegaMapper \cite{MegaMapper} or MSE \cite{MSE}, though significantly sparser than one might expect from future $21\,$cm experiments \cite{Slosar19a}.  We will highlight the results from the denser sample --- which we will call the fiducial sample throughout --- since it is less noisy, but the results are qualitatively similar for the lower density sample.

\section{Feature models}
\label{sec:models}

We will consider two broad classes of ``features'' in the linear theory power spectrum.  The first will arise in the very early Universe (primordial features), for example when the perturbations were originally generated by inflation.  The second will be imprinted after inflation but at much earlier times than the epoch of the observations.  Let us take each in turn.

\subsection{Primordial features}

\begin{figure}
    \centering
    \resizebox{\columnwidth}{!}{\includegraphics{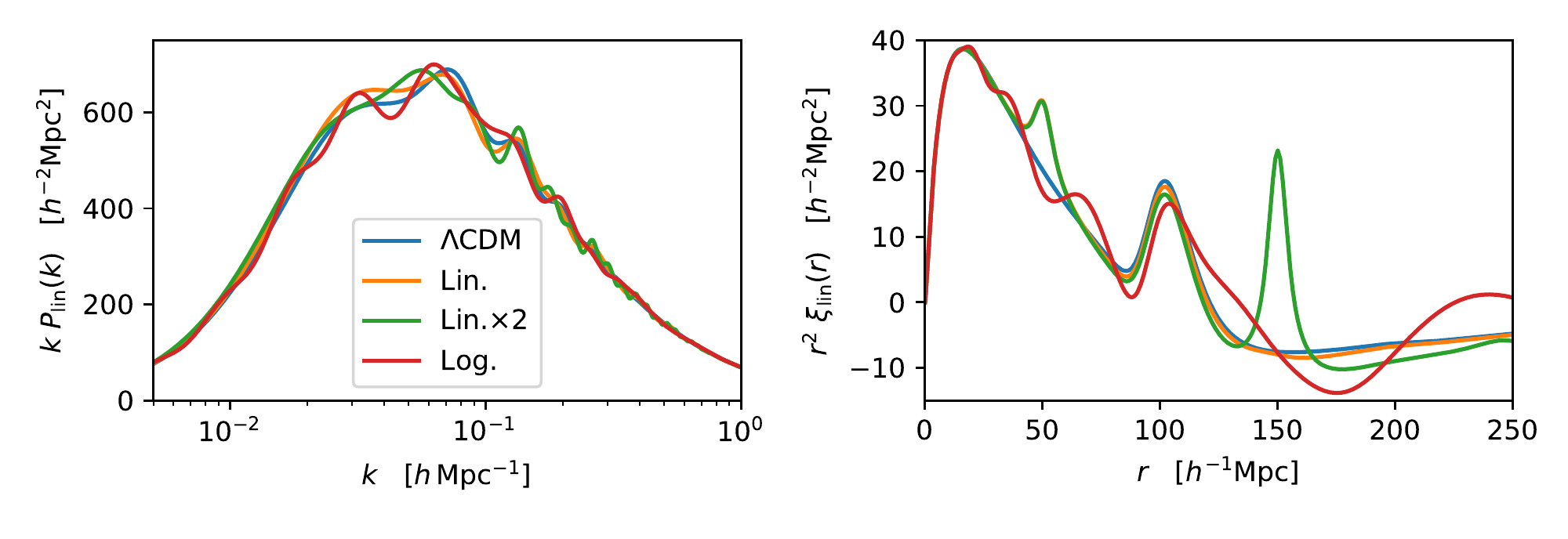}}
    \caption{The linear theory power spectra (left) and correlation functions (right) for our fiducial $\Lambda$CDM model and models with primordial features superposed.  The ``Lin.'' model has sinusoidal oscillations linear in $k$ (Eq.~\ref{eqn:lin}), the ``Lin.$\times 2$'' model has sinusoidal oscillations linear in $k$ with two frequencies ($\omega_1=50\,h^{-1}$Mpc and $\omega_2=150\,h^{-1}$Mpc) and the ``Log.'' model has sinusoidal oscillations in $\ln k$ (Eq.~\ref{eqn:log}).  See text for further details.}
    \label{fig:primordial}
\end{figure}

We investigate several phenomenological models of primordial features, chosen to illustrate various issues and highlight results, rather than models based on fundamental physics calculations.  Specifically we follow the recent literature in considering two types of oscillations that are superposed upon the linear theory power spectrum computed for $\Lambda$CDM,
\begin{equation}
    P_{\rm lin}(k) = P_{\Lambda\mathrm{CDM}}(k)
    \left\{ 1 + A\sin\left(\omega k\right)\exp\left[-\frac{\left(kr_d\right)^2}{2}\right]\right\}
    \qquad \mathrm{(linear)}
\label{eqn:lin}
\end{equation}
and
\begin{equation}
    P_{\rm log}(k) = P_{\Lambda\mathrm{CDM}}(k)
    \left\{ 1 + A\sin\left(\omega\ln\frac{k}{k_\star}\right)\exp\left[-\frac{\left(kr_d\right)^2}{2}\right]\right\}
    \qquad \mathrm{(logarithmic)}
\label{eqn:log}
\end{equation}
with $A=0.05$, $k_\star=0.05\,h\,{\rm Mpc}^{-1}$ and $r_d=2.5\,h^{-1}$Mpc.  The first class of models have oscillations linear in $k$, often termed ``sharp features'', and tend to arise if the inflaton temporarily departs from its slow roll (attractor) evolution.  The second class, with oscillations in $\ln k$, are also termed ``resonant features'' \cite{Slosar19b}.
Compared to earlier work we have chosen a particular phase for the linear oscillations so that the modification tends to zero at low $k$ and damped the models at high $k$ with a Gaussian.  The high $k$ damping more closely reproduces the models based on features in the inflationary potential which tend to produce band-limited oscillations (e.g.\ refs.\ \cite{Adams01}), and also ensures that our simulations are properly, numerically resolving the features.  While models with 5 per cent oscillations such as these are observationally disfavored \cite{PlanckInf18,Beutler19}, using larger amplitude oscillations provides higher signal to noise in our simulations and a more stringent test of the modeling formalism.  For the linear model we choose $\omega=50\,h^{-1}$Mpc, to emphasize non-linear evolution of the feature compared to the BAO, while for the logarithmic model we take $\omega=10$, which ensures we resolve the oscillations well with our $2.5\,h^{-1}$Gpc box.  The linear theory power spectra and correlation functions, extrapolated to $z=0$, are shown in Fig.~\ref{fig:primordial}.

One of the advantages of the linear model is that it can be thought of as a single mode in a Fourier decomposition of a more general class of features.  Since all of our models are built upon a $\Lambda$CDM template, a second feature (due to baryon acoustic oscillations in the recombination-era Universe \cite{Dodelson03}) is always present.  However, in order to gauge how well we can model non-linear evolution, bias and redshift-space distortions in the presence of multiple frequencies we also generate a linear model with two sine modes of frequencies $\omega_1=50\,h^{-1}$Mpc and $\omega_2=150\,h^{-1}$Mpc (Fig.~\ref{fig:primordial}).  Each mode has the same damping and amplitude as for the ``linear'' model above.

\subsection{Induced features}

\begin{figure}
    \centering
    \resizebox{\columnwidth}{!}{\includegraphics{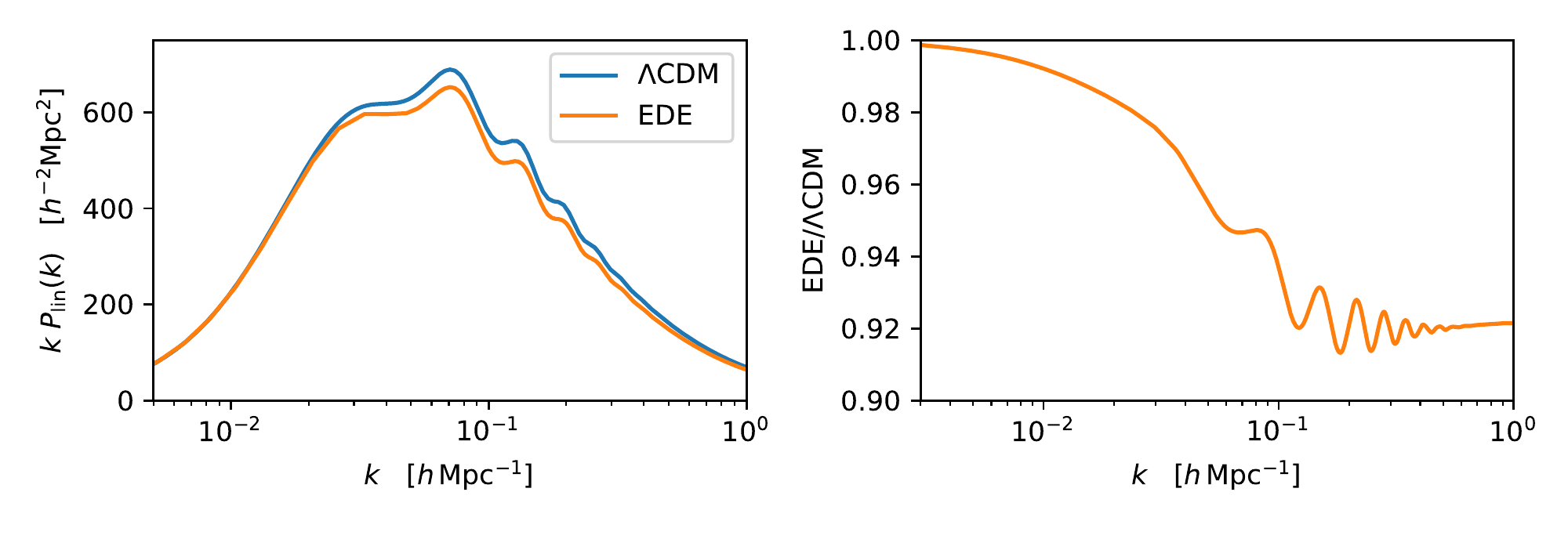}}
    \caption{The linear theory power spectra (left) and their ratio (right) for our fiducial $\Lambda$CDM model and models with features induced by a period of early dark energy (EDE) at $z\simeq 10^4$.  See text for further details.}
    \label{fig:induced}
\end{figure}

As a second class of features we consider changes to the matter power spectrum that arise due to non-standard expansion histories.  Since the growth of structure is damped by Hubble expansion, long periods where unclustered species dominate the expansion lead to suppression of large-scale structure that can be detected by comparing early- and late-time measures of the fluctuation amplitude.  A precise measurement of the power spectrum shape can also be used to place constraints on (or detect) short-term deviations from matter or radiation domination, since such periods will change the shape of the power spectrum due to differential growth of modes.  Recently this reasoning was used to place constraints on early dark energy (EDE) models which contribute to the expansion near recombination \cite{Hill20,Ivanov20,DAmico20b,Klypin20}.  The point is more general however, and an accurate measurement of the power spectrum constrains deviations in the expansion history over a broad range of redshifts\footnote{While we do not consider it here, features can also be introduced by interactions between or among particle species}.

As an exploration of this class of effects we consider scalar-field based models of Early Dark Energy (EDE) wherein the impact of EDE is localized to time significantly before those probed observationally. EDE models are a timely example as they have been the subject of much recent interest as a potential way to resolve discordances between $\Lambda$CDM analyses of various data sets \cite{Smith19,Smith20,Hill20}. In fact we use the modification of CLASS \cite{CLASS} by the authors of ref.~\cite{Hill20} and consider a model where the contribution from EDE peaks at $z\simeq 10^4$ with peak fractional contribution (to $\rho$) of 10 per cent.

The linear theory power spectrum for this model is compared to our fiducial $\Lambda$CDM model in Fig.~\ref{fig:induced}, with the right panel showing the ratio to better highlight the change in shape induced by the non-standard expansion history.  The position and amplitude of the feature in the right panel of Fig.~\ref{fig:induced} are set primarily by the redshift at which the EDE becomes non-negligible and the fraction of the energy density in dark energy (respectively).

\section{Peturbative model}
\label{sec:PT}

\subsection{Overview of Previous Work}

The effect of non-linearities and long-wavelength modes (IR-resummation) on oscillatory features in the power spectrum has long been studied in the context of BAO.  A large body of work shows that these features can be accurately modeled in Lagrangian \cite{Bha96,Mat08a,Mat08b,ESW07,PWC09,Noh09,CLPT,TasZal12,McCSza12,White14,Vlah16,McQuinn16} and Eulerian \cite{Crocce08,SenZal15,Schmittfull15,Baldauf15,Blas16,Seo16,Ding17,Hikage17,Peloso17,Ivanov18} perturbation theory.

A perturbative analysis of features based on the Eulerian framework has been done recently \cite{Beutler19, Vasudevan19} (see also ref.~\cite{Ballardini20}), in which the authors studied the effects of long wavelength modes on the primordial features of types given by Eqs~\eqref{eqn:lin} and \eqref{eqn:log}.  An important finding of these studies is that higher-loop corrections give rise to $\mathcal{O}(1)$ modifications to the features for a wide range of parameters and wavenumbers of interest. This leads one to consider resumming these $\mathcal{O}(1)$ contributions, arising from the large displacements, as is done in the case of BAO (see above).

For a general, oscillatory power spectrum component, refs.~\cite{Beutler19, Vasudevan19} show that the long modes' effect on the one-loop contribution can be computed as 
\begin{equation}
P^{\rm w}_{\rm 1-loop}(k) = \frac{1}{2} \int^\Lambda \frac{d^3p}{(2\pi)^3} ~ \frac{(\bp \cdot \bk)^2}{p^4} P^{\rm nw}(k)
\Big[ P^{\rm w} (|\bk + \bp|) + P^{\rm w}  (|\bk - \bp|) - 2 P^{\rm w}(k) \Big],
\end{equation}
where $\Lambda$ is a cut-off scale such that $p<\Lambda \sim k$.  Taylor expanding the first two components in the brackets of integrand in $q$ and formally integrating gives a one-loop contributions of the form
\begin{equation}
  P^{\rm w}_{\rm 1-loop, X}(k) = - \frac{1}{2} k^2 \Sigma_X^2 P^{\rm w}(k),
\end{equation}
where $\Sigma_X$ is the effective displacement dispersion at a point and $X$ labels either the linear, or logarithmic shapes given in Eqs~\eqref{eqn:lin} and \eqref{eqn:log}.
The dispersion can be well approximated by
\begin{equation}
\Sigma^2_{X} = \frac{1}{3\pi^2} \int_0^\Lambda dp ~ \left[ 1 - j_0 \left( \omega_X p \right) +2 j_2 \left( \omega_X p \right) \right] P^{\rm nw}(p),
\label{eqn:ept_sigma}
\end{equation}
where for linear\footnote{These derivations always assume the $r_d \to 0$ limit.} shapes (Eq.~\eqref{eqn:lin}) we have $\omega_X = \omega_{\rm lin}$, while for logarithmic shapes we have 
$\omega_X= \omega_{\rm log}/k$.  The latter makes a further approximation that, due to the shape of $P^{\rm nw}$, the $\Sigma_{\rm log}$ integral has most of its contributions from the $p\ll k$ part of the integral.
In the case of linear oscillations the above expression corresponds to the results obtained in BAO studies and their IR-resummations.

Following the results of ref.~\cite{Vasudevan19}, the total non-linear matter power spectrum at one-loop can be obtained as 
\begin{equation}
  P_{{\rm 1-loop}, X}(k) = P^{\rm nw}(k)_L +  \left(  1 + \frac{1}{2} k^2 \Sigma_X^2 \right) e^{-\frac{1}{2} k^2 \Sigma_X^2} P^{\rm w}(k)_L 
  + P^{\rm nw}_{\rm 1-loop} \left[ P^{\rm nw}_L + e^{-\frac{1}{2} k^2 \Sigma_X^2} P^{\rm w}_L \right] (k).
  \label{eqn:ept_damping}
\end{equation}
This expression can readily be extended to the power spectrum for biased tracers, since the above IR-resummation procedure remains unchanged.  Moreover we see that if more than a single distinct feature is present in the power spectrum (as is the case if one studies e.g.\  BAO and some other feature), the above expression simply obtains additive $P^{\rm w}$ contributions, given that, to a very good approximation, we can neglect the cross-correlation  contributions of different wiggle components (e.g.\ $P^{\rm w}_{\rm lin}\times P^{\rm w}_{\rm BAO}$). For further details on these results we refer a reader to refs \cite{Beutler19, Vasudevan19}.  In what follows we shall show how the IR resummation is naturally handled in LPT, and how this leads to a different way of obtaining an Eulerian resummation procedure.

\begin{figure}
    \centering
    \includegraphics[width=\textwidth]{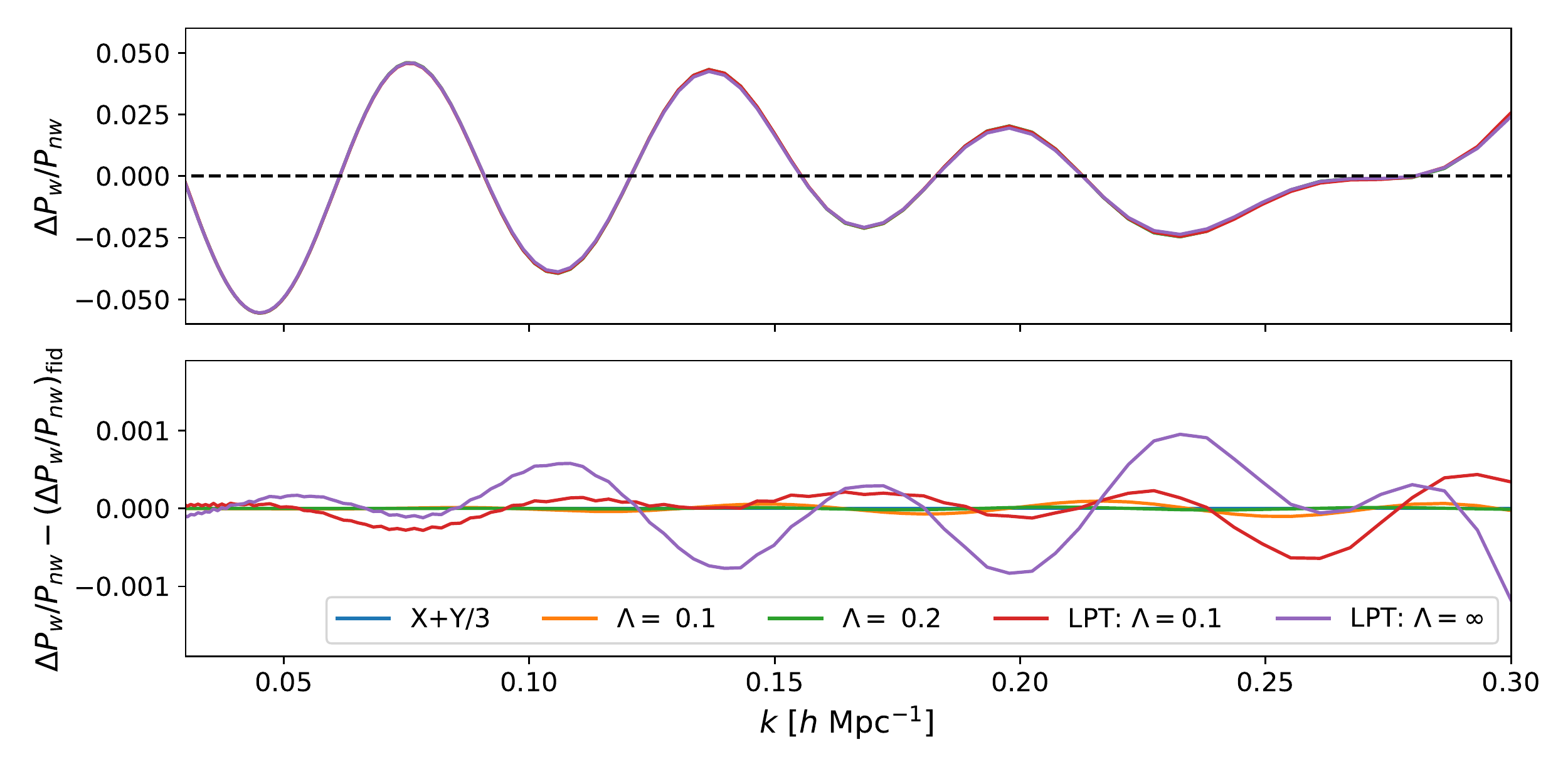}
    \caption{Comparison of the oscillatory components of the real-space power spectrum for our fiducial halo sample at $z = 1$ as predicted by 1-loop EPT and LPT for a range of IR-resummation choices in the $\Lambda$CDM cosmology. All choices are in excellent numerical agreement -- the EPT schemes are all within $10^{-4}$ of the total broadband power of each other and differ at the $10^{-3}$ level with the LPT prediction. The latter number lets us place a minimum theoretical error on predictions for feature amplitude.}
    \label{fig:ir_compare}
\end{figure}

\subsection{Lagrangian IR Resummation and Connection to Earlier Approaches}

In contrast to the above, within the Lagrangian framework (LPT) IR resummation can be naturally incorporated by exponentiating long-wavelength displacements \cite{Mat08a,CLPT,Vlah16}. Within $\Lambda$CDM this is nearly equivalent to simply exponentiating the linear displacements\edit{, since in such cosmologies the variance of displacements in linear theory due to modes at high $k$ is relatively small while those at low $k$ are approximately linear. In this regime} the matter power spectrum is given by
\begin{equation}
    P_{m} = \dtq e^{i \bk \cdot \bq - \frac{1}{2} k_i k_j A^{\rm lin}_{ij}}\ \Big\{1 - \frac{1}{2} k_i k_j A^{\rm loop}_{ij} + \frac{i}{6} k_i k_j k_k W_{ijk} + ... \Big\},
\end{equation}
where $A_{ij}$ and $W_{ijk}$ are $n$-point statistics of pairwise displacements $\Delta = \bPsi_1 - \bPsi_2$ with separation $\bq = \bq_1 - \bq_2$. 
\edit{In the expression above, in principle only the long modes of the linear displacement field should be  resummed. This can be accomplished by splitting the exponentiated $A_{ij}$ in the above equation with the same kind of cutoff $\Lambda$ as in Equation~\ref{eqn:ept_sigma} and Taylor-expanding the short-wavelength component\footnote{In practice we use a Gaussian cutoff. See also Section 4.3 in \cite{Chen20}.}. However, the cumulative effect of introducing such an explicit IR scale, $\Lambda$, on any wiggle shapes (including BAO) is quite small, constituting less then the 0.1\% difference, as shown in Figure \ref{fig:ir_compare}.
This approximation is of course true as long as the resummed displacement dispersion does not receive large contributions from small scales, which is the case for $\Lambda$CDM-like spectra.}

Within LPT, the Eulerian resummation can be recovered as a saddle-point approximation; briefly, for an input linear power spectrum with smooth and wiggly components $P_{\rm lin} = P_{nw} + \Delta P_X$, the latter can be expanded out of the exponent, where its configuration-space feature at some characteristic separation $q_X$\footnote{In general the scale depends upon the first derivative of the argument of the sine. For logarithmic wiggles the Eulerian treatment is equivalent to making the approximation
\begin{equation}
    \sin \Big( \omega \ln \frac{k}{k_\ast} \Big) \approx \sin \Big( \frac{\omega}{k_0} (k - k_0) + \phi_0 \Big),
\end{equation}
which at each $k$ picks out the characteristic scale $q_X(k) = \omega/k$. This approximation works increasingly well the larger $\omega$ is (see Appendix \ref{app:saddle}).} will pick out a nonlinear smoothing
\begin{equation}
    e^{-\frac{1}{2} k^2 \Sigma_X^2} = \Big \langle \exp\Big\{ -\frac{1}{2} k_i k_j A^{nw,\rm lin}_{ij}(\bq) \Big\}_{|\bq|=q_X} \Big \rangle.
    \label{eqn:lpt_damping}
\end{equation}
This recovers the $\Lambda \rightarrow \infty$ limit of Eq.~\ref{eqn:ept_sigma}. The brackets in the above equation refer to the angular average of the exponent in the space of $\bq$'s.  Decomposing $A_{ij} = X(q) \delta_{ij} + Y(q) \hat{q}_i \hat{q}_j$ into scalar components $X$ and $Y$, the most straightforward approach is to take the average to correspond to $X + Y/3$, though we note for example that Eq.~\ref{eqn:ept_sigma} corresponds to taking $X + Y$. If there are multiple oscillatory components, i.e.\ a superposition of sinusoidal components as in ``Lin.x2'' (Fig.~\ref{fig:primordial}), then as long as each component is perturbatively small the above argument can be applied independently to determine the saddle-point $q_X$ and damping factor $\Sigma^2_X$ of each feature as argued for EPT in the discussion below Eq.~\ref{eqn:ept_damping}.

Based on the Eulerian and Lagrangian discussions above it may seem like there exists an over-abundance of possible IR resummation schemes. This is not, however, a matter of great concern since it turns out all of these schemes behave quantitatively similarly, at least within roughly $\Lambda$CDM cosmologies. \edit{For example, as noted above the difference between setting $\Lambda$ to be some fraction of $k$ and letting $\Lambda \rightarrow \infty$ as in LPT will be small in such cosmologies due to the relative smallness of linear displacements at high $k$.} Figure~\ref{fig:ir_compare} shows a comparison of these schemes for our fiducial $\Lambda$CDM cosmology at $z = 1$, with bias parameters taken from our fiducial halo sample. \edit{For clarity of presentation we have isolated the oscillatory signal by subtracting a rough broadband computed using a Savitsky-Golay filter, then supplementing each theory curve with a quartic polynomial in $k$ such that curves identical modulo such a quartic contribution will be coincident}\footnote{In this and other plots of the nonlinear oscillation signal below, we remove broadband differences by fitting a quartic polynomial in $k$ to the differences between plotted curves.}. Taking $ \Sigma^2_X = X + Y/3$ derived from Eq.~\ref{eqn:lpt_damping} as our baseline (blue curve), we see that the differences between this choice and the conventional EPT dampings with $\Lambda = 0.1$, $0.2\,h{\rm Mpc}^{-1}$ are extremely small and at the level of $10^{-4}$ when compared to the (linear theory) broadband power. This is because much of the numerical difference between the ``linear'' $e^{-\frac{1}{2}k^2 \Sigma_X^2} \Delta P_w$ is ameliorated by correctly accounting for damping effects at one-loop level in Eq.~\ref{eqn:ept_damping}.  On the other hand, the differences between these schemes and a direct LPT calculation in which the linear displacements are fully resummed is larger, though still in excellent numerical agreement, at about the $10^{-3}$ level\edit{, while an LPT calculation with $\Lambda = 0.1 h$ Mpc$^{-1}$ lies in between}. Finally, let us note that while the (Gaussian) statistical error on power spectrum measurements scale as amplitude of the total power, the theory error discussed above scales as the amplitude of the wiggles only; for example, while it is at $10^{-3}$ of the total power for a $5\%$ feature (BAO), non-BAO primordial features bounded at the $1\%$ level using BOSS and Planck data by ref.~\cite{Beutler19}, it will be at the $2 \times 10^{-4}$ level at the same redshift.

From the above comparison, we can conclude that (1) the disagreement between LPT and EPT lets us set a minimum theoretical error on theoretical predictions for feature ampltidue at about $0.1\%$ and (2) that the difference between the various Eulerian IR resummation schemes and their predictions for $\Sigma_X^2$ are small compared to this theoretical error, such that we can be reasonably cavalier when choosing between them. Indeed, as $k$ approaches the nonlinear scale we should expect that oscillatory signals from beyond-one-loop contributions will become increasingly prominent compared to the amplitude of the linear oscillations, dwarfing the theoretical differences highlighted above in the same way that the scale of the bottom panel of Figure~\ref{fig:ir_compare} is much smaller than that of the top panel. Given that the one-loop EPT-LPT difference is only a few percent of the wiggle amplitude in $\Lambda$CDM---5 percent compare to $0.1$ percent---even at the edge of our perturbative reach at $k = 0.2 h$ Mpc$^{-1}$, the above comparison suggests that searches for primordial features should focus on exploring higher redshifts and volumes.

\subsection{Redshift-Space Distortions}

Since next-generation galaxy redshift surveys will be the natural hunting ground for primordial features, the focus of this paper is to extend the modeling of primordial features in previous works to redshift space. We use 1-loop Lagrangian and Eulerian perturbation theory (LPT and EPT) described in the previous subsections to model both the real- and redshift-space power spectra of biased tracers (i.e.\ halos in our context) \edit{within the plane parallel approximation} as described in detail in ref.~\cite{Chen20}.  The redshift-space power spectrum, $P_s(k,\mu)$, is evaluated as an expansion in the line-of-sight wavenumber, $k_\parallel=k\mu$, multiplying $n^{\rm th}$-order pairwise velocity spectra, such that
\begin{equation}
    P_s(k,\mu) = \sum_{n=0}^\infty \frac{(i k \mu)^n}{n!} \tilde{\Xi}^{(n)}_{\parallel}(k,\mu).
\end{equation}
The model includes Lagrangian and Eulerian third-order bias expansions along with counter terms and stochastic terms to account for small-scale physics that we do not explicitly model.  Our fiducial model includes terms up to second order in the velocity expansion, but employ an ansatz for the third moment which is shown to be highly accurate for $\Lambda$CDM-like models \cite{Chen20}. As discussed in ref.~\cite{Chen20}, while a finite-order expansion in the velocity moments $\tilde{\Xi}$ necessarily omits feature damping due to bulk velocities that would be included in a ``complete'' IR resummation scheme, particularly at high $\mu$, we will show that it is nonetheless sufficiently accurate for upcoming galaxy surveys like DESI; however, for completeness we also include comparisons to (1) the Gaussian streaming model and (2) one-loop EPT wherein both bulk displacements and velocities are resummed where appropriate. We set the third order Lagrangian bias to zero since its effects are subdominant for the halos and scales of interest \cite{Lazeyras16,Abidi18}.

\section{Results}
\label{sec:results}

\begin{figure}
    \centering
    \resizebox{\columnwidth}{!}{\includegraphics{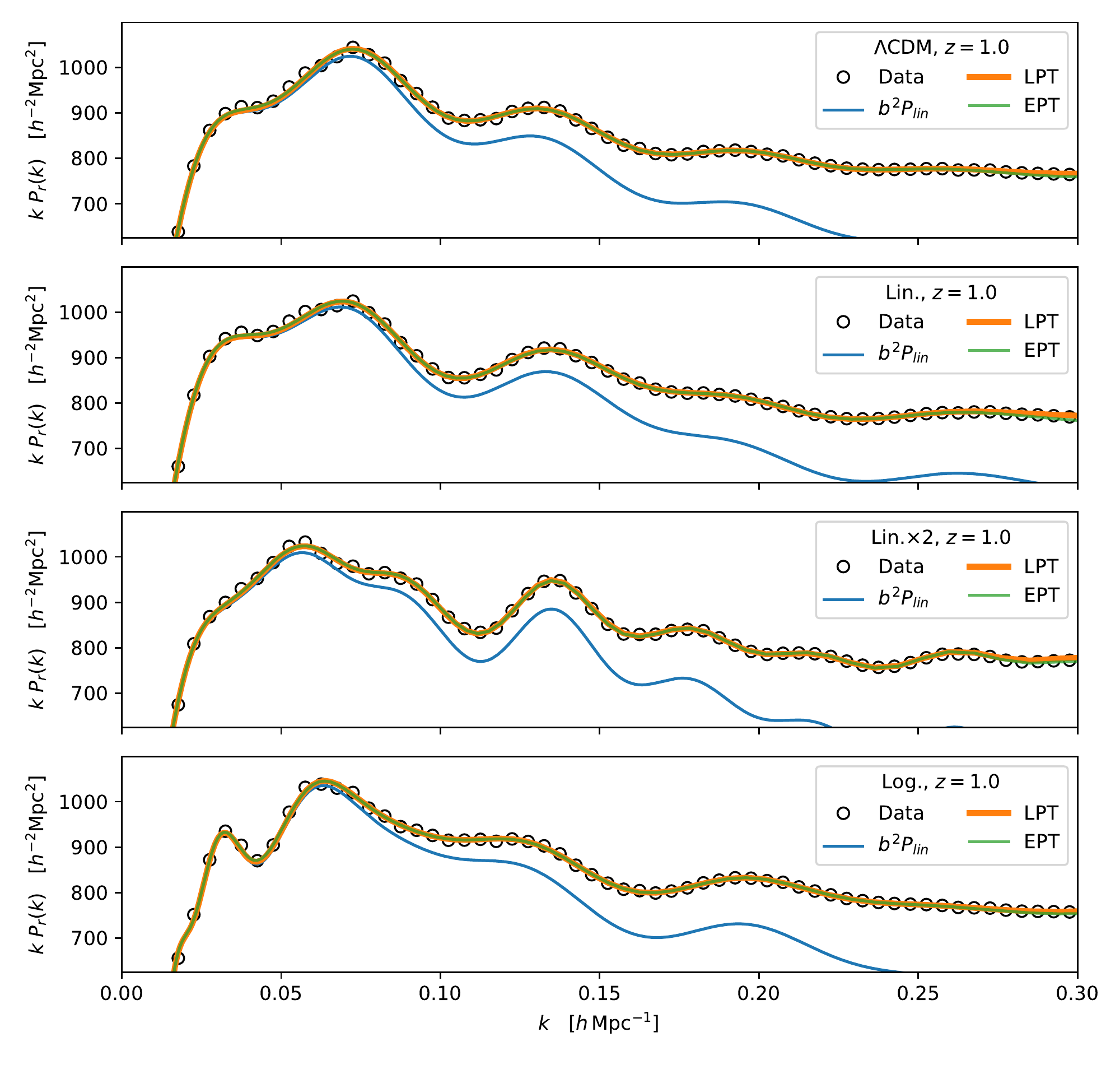}}
    \caption{The real-space, halo power spectra from our simulations at $z=1$ and model fits.  We show results for the $\bar{n}=10^{-3}\,h^3\,{\rm Mpc}^{-3}$ sample, since it has lower shot noise, but results for the sparser sample are qualitatively similar.  The open, black circles show the average of $P(k)$ over the 4 boxes. The orange and green lines (which are almost on top of each other) show the best-fit LPT and EPT models while the blue line shows linear theory with the same large-scale bias as the EPT models.}
    \label{fig:real}
\end{figure}

\begin{figure}
    \centering
    \includegraphics[width=\textwidth]{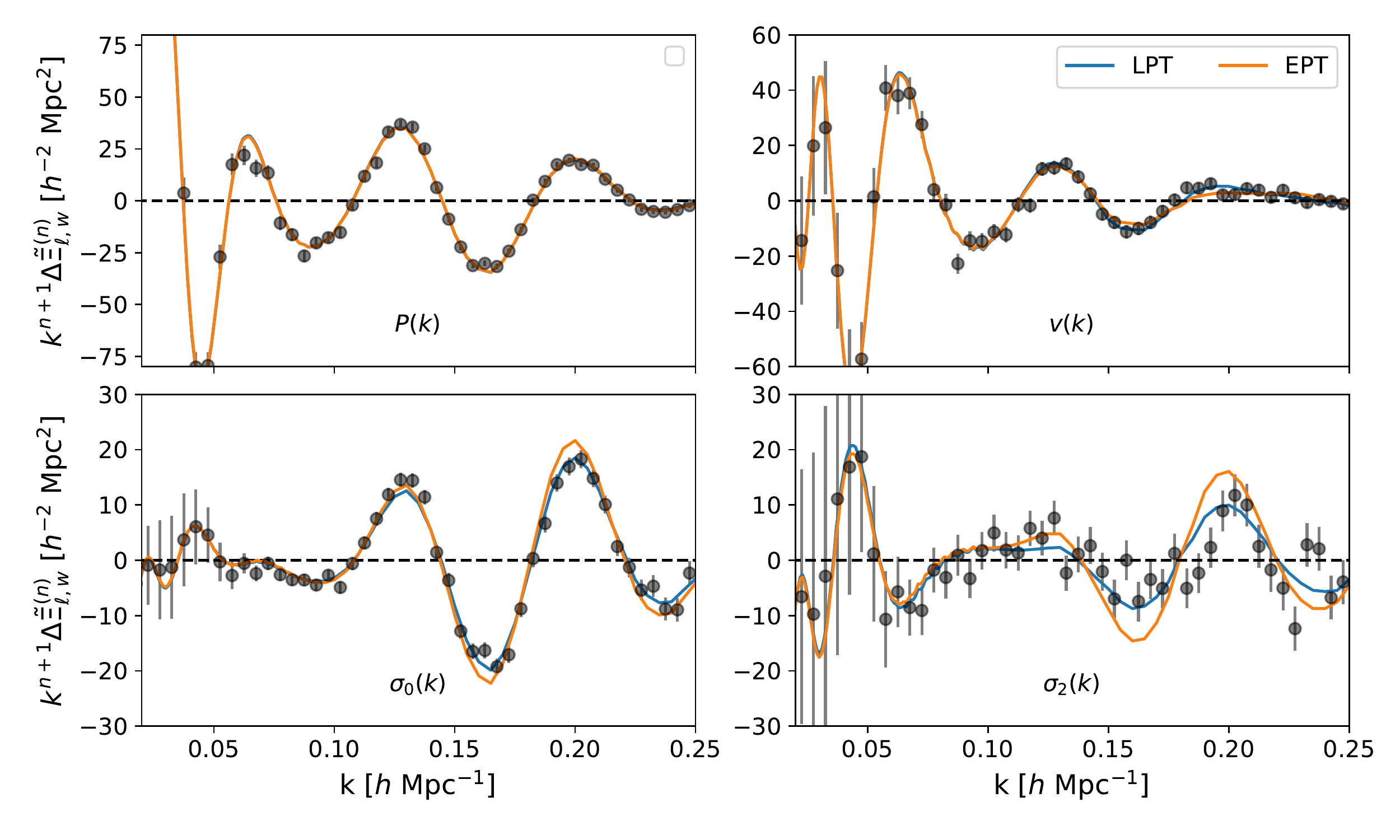}
    \caption{Broadband-subtracted pairwise-velocity moments in real space for our fiducial halo sample with the ``Lin.$\times 2$'' linear power spectrum compared to predictions from LPT and EPT. As in basic $\Lambda$CDM models, there is excellent quantitative agreement between LPT and EPT in the zeroth and first moments, while in the second moment EPT slightly underpredicts the damping of 1-loop wiggles prominent at higher $k$.
    }
    \label{fig:wiggles_vel}
\end{figure}

To see how well our perturbative models predict the non-linear evolution and redshift-space distortions in models with primordial or induced features we compare to our N-body results.  For each model we fit to the average of the 6 simulations.  The fits are done using the model ``out of the box'', i.e.\ without tweaking or adjusting any settings, letting only bias parameters and effective corrections float, and assuming the correct cosmology and linear theory power spectrum.

Figure \ref{fig:real} shows an example of our fits to the real-space halo spectra for the primordial feature models.  The real-space spectra are not directly observable (except in projection, which will tend to wash out the features) but serve to show that the model is able to predict the underlying clustering well.  We focus on the middle redshift ($z=1$) and highest number density ($\bar{n}=10^{-3}\,h^3\,{\rm Mpc}^{-3})$ sample since this has the smallest error bars.  The agreement between both models and N-body data is excellent over the entire range of quasi-linear scales and into the regime where shot-noise begins to dominate the spectra.  While not shown, we have checked that the more biased sample and other redshift slices show similar levels of agreement.

In addition to the real-space power spectrum we have also measured the first two, real-space pairwise velocity moments of our halo samples and compared them to perturbation theory. These velocity statistics inform the angular structure of the redshift-space power spectrum (which we expect will eventually provide our tightest observational constraints on features) in addition to being well-defined observables in their own right, and extracting them individually gives us a closer look at oscillatory features that only become prominent at high $\mu$. Figure~\ref{fig:wiggles_vel} shows our best-fit LPT and EPT models for the model with two linear oscillations (Lin.$\times 2$), again with broadband shapes subtracted to isolate the oscillatory components. Much as in $\Lambda$CDM models \cite{Chen20} there is excellent agreement between LPT and EPT for the real-space power spectrum ($P$), and the pairwise velocity ($v$), while EPT tends to slightly underpredict the damping for the second moment ($\sigma_\ell$), especially as the one-loop oscillations become prominent at $k > 0.1\, h\text{Mpc}^{-1}$.  These differences are, however, small and the models predict almost identical power spectrum wedges. Nonetheless, they are helpful in informing our theoretical error budget when searchiing for oscillations close to the line-of-sight.

As Figs.\ \ref{fig:real} and \ref{fig:wiggles_vel} make clear the Eulerian and Lagrangian descriptions provide almost identical performance over the range of scales where we expect perturbation theory to be valid.  We find this persists even for the redshift-space spectra, and so to avoid clutter we shall show only the Lagrangian model in the following figures.

\begin{figure}
    \centering
    \resizebox{\columnwidth}{!}{\includegraphics{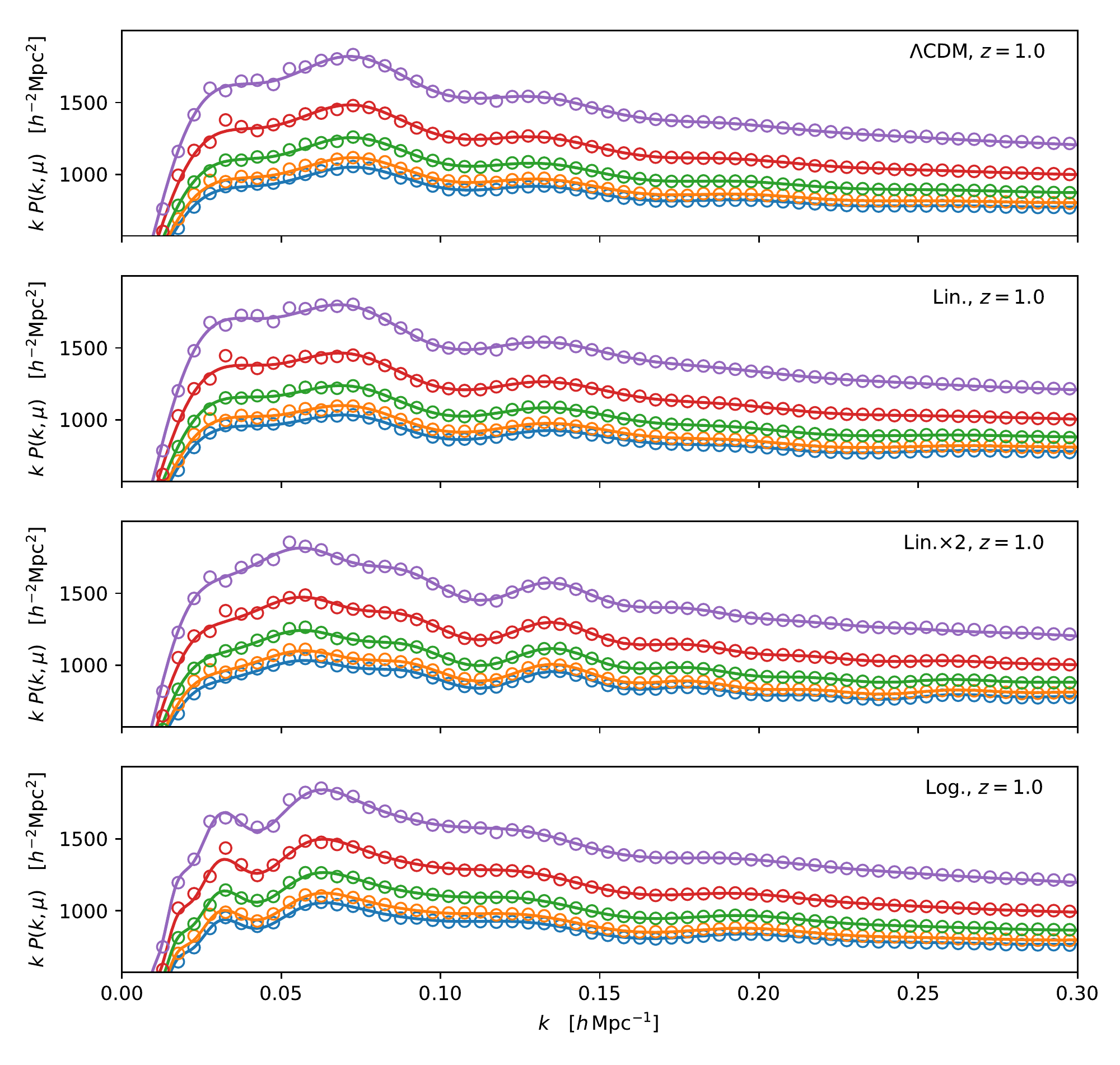}}
    \caption{The redshift-space, halo power spectrum wedges from our simulations at $z=1$ and model fits.  We show results for the $\bar{n}=10^{-3}\,h^3\,{\rm Mpc}^{-3}$ sample, since it has lower shot noise, but results for the sparser sample are qualitatively similar.  The open circles show the average of $P(k,\mu)$ for $\mu=0.1$, $0.3$, ..., $0.9$ (colors, bottom to top), the solid lines show the best-fit LPT model.}
    \label{fig:wedge}
\end{figure}

\begin{figure}
    \centering
    \includegraphics[width=0.95\textwidth]{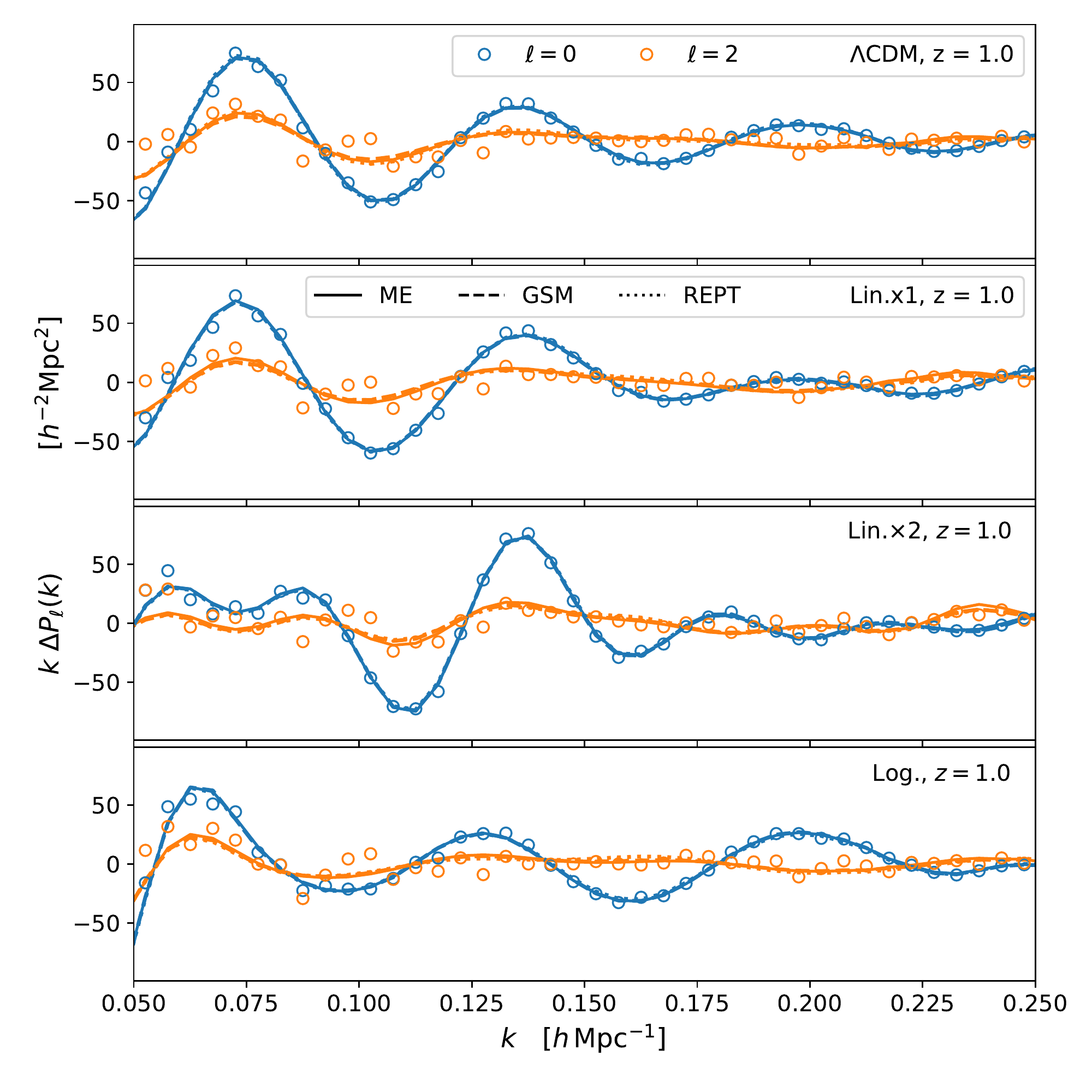}
    \caption{Predictions using the LPT moment expansion, Gaussian streaming model and IR-resummed EPT (REPT) for the oscillatory components of the redshift-space power spectrum monopole and quadrupole at $z = 1$ for the $\bar{n}=10^{-3}\,h^3\,{\rm Mpc}^{-3}$ sample. The LPT and REPT models are in excellent agreement, especially compared to the scatter of the N-body data to which they were independently fit.}
    \label{fig:method_pells}
\end{figure}

Figure \ref{fig:wedge} compares the theory prediction (LPT moment expansion) and N-body data for the anisotropic power spectrum wedges $P(k,\mu)$ in redshift space. The theory and data are in excellent agreement over a large range of scales and LOS angles $\mu$, though we note that the shot noise $\sim \bar{n}^{-1}$ plays an increasingly dominant role at the highest $k$'s shown.  We focus on the redshift-space wedges, which are independent in linear theory, but have checked that the theory returns an equivalently good fit to the smooth and oscillatory components of the multipoles.

As an additional check on possible theory differences, Fig.~\ref{fig:method_pells} shows the predictions for the oscillatory components of $P_\ell(k)$ (i.e.\ with the broadband subtracted) for our fiducial halo sample using our LPT moment expansion model, Gaussian streaming model and resummed Eulerian perturbation theory. Despite the Lagrangian and Eulerian models being fit separately to the wedges, i.e.\  $P(k,\mu)$ data, all three models are in excellent agreement with each other and the N-body data for the multipoles, $P_\ell(k)$.  The agreement is particularly impressive in comparison to the scatter in the N-body data, which are themselves tighter than expected for upcoming surveys with a total simulated volume of $>90\,h^{-3}{\rm Gpc}^3$. In order to demonstrate their equivalence at low $k$, we have matched the countererms between the moment expansion and Gaussian streaming model predictions (Appendix~\ref{app:gsm}), but note that at the highest $k$'s shown the GSM predictions for the quadrupole are in fact in slightly better agreement with the EPT prediction, suggesting that differences there are driven by the incomplete IR resummation of bulk velocities in the moment expansion (see ref.~\cite{Chen20} for further discussion).

\begin{figure}
    \centering
    \resizebox{\columnwidth}{!}{\includegraphics{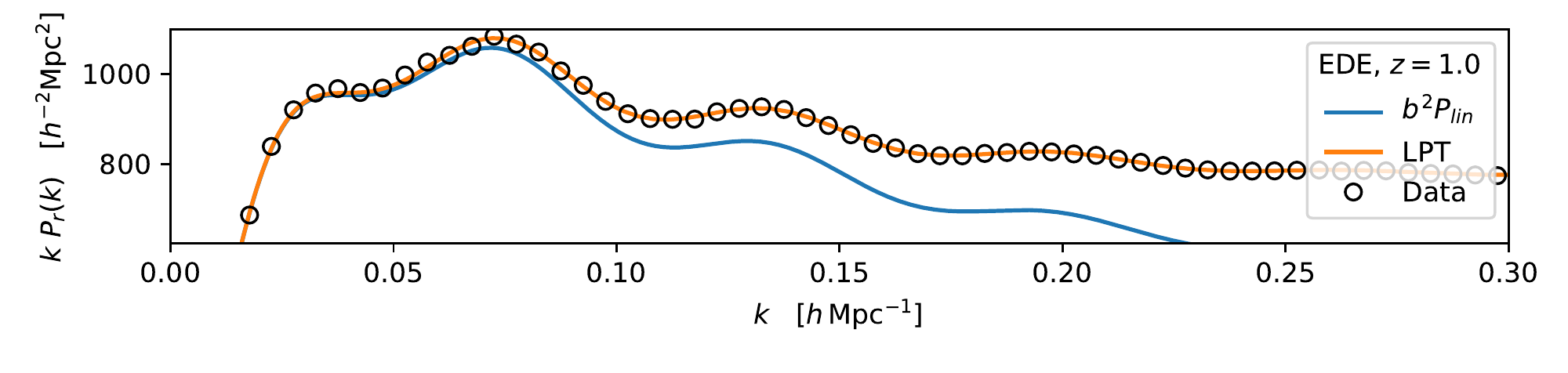}}
    \resizebox{\columnwidth}{!}{\includegraphics{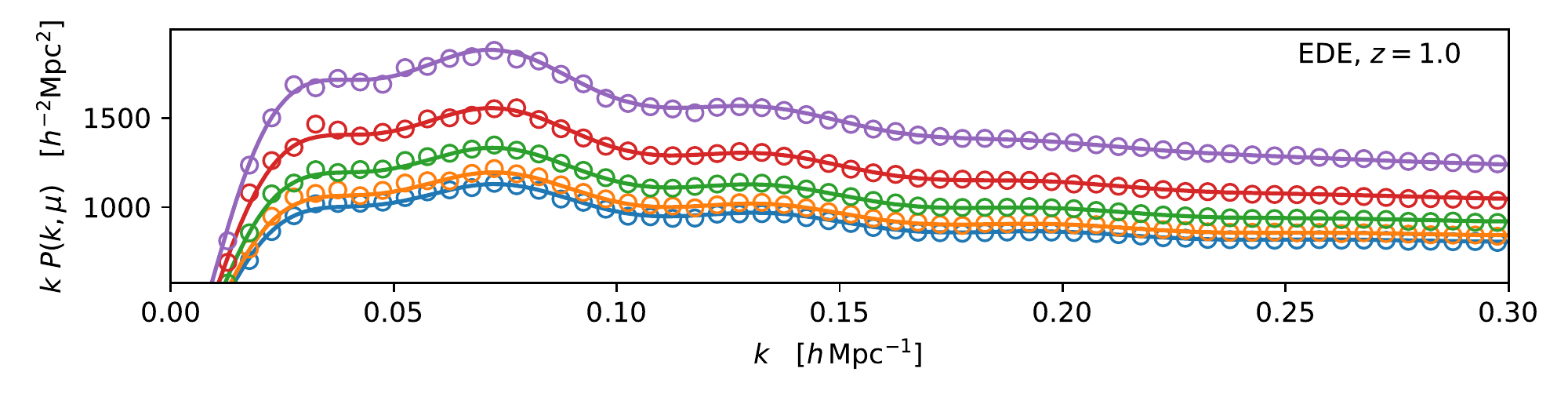}}
    \caption{As for Figs.~\ref{fig:real} and \ref{fig:wedge} but for the model with an ``induced feature'' (see text and Fig.~\ref{fig:induced}).}
    \label{fig:ind}
\end{figure}

Finally, Fig.~\ref{fig:ind} shows similar fits to the model with 10 per cent early dark energy at $z\simeq 10^4$.  Again the \edit{fiducial LPT} model provides an excellent fit to the N-body data in both real and redshift space on quasi-linear scales, indicating that these scales can be used to constrain the amount of unclustered dark energy contributing to the expansion when such modes entered the horizon. \edit{A similar level of fit is obtained using other schemes such as resummed EPT, though we have not shown them for sake of brevity.} Future surveys, probing large volumes at high redshift, should be able to constrain the expansion history via its effect on growth over a broad range of cosmic history.

\section{Conclusions}
\label{sec:conclusions}

We have investigated how well 1-loop perturbation theory, both Eulerian and Lagrangian, can model the redshift-space power spectra of biased tracers such as dark matter halos in models with either primordial or induced features.  By comparing the models of ref.~\cite{Chen20} to clustering statistics measured from a series of large N-body simulations, we have shown that bias, non-linear evolution and redshift-space distortions can all be accurately accounted for by existing perturbative models with no need for any tuning or modification.  This is the first demonstration that such perturbative models can fit the redshift-space clustering of biased tracers in such theories to per cent level precision on quasi-linear scales.

We compare three primordial feature models (Fig.~\ref{fig:primordial}) and one model with an ``induced'' feature imprinted by an epoch of early dark energy (Fig.~\ref{fig:induced}).
We investigated two different schemes for modeling the dynamics (Eulerian and Lagrangian), several different IR resummation schemes, and different methods for including redshift-space distortions (a moment expansion and a cumulant expansion: the Gaussian Streaming Model).  In all cases we find excellent agreement between the different theories and the N-body simulations.  Figure \ref{fig:ir_compare} shows that different methods of performing IR resummation give extremely similar predictions for the power spectrum.  Figures \ref{fig:real} and \ref{fig:wiggles_vel} show that both LPT and EPT predict the real-space density and velocity statistics measured in our N-body simulations well to $k\simeq 0.25\,h\,{\rm Mpc}^{-1}$.   With these ingredients, Fig.~\ref{fig:wedge} shows that the moment expansion accurately describes the redshift-space, power spectrum wedges, $P(k,\mu)$ with Fig.~\ref{fig:method_pells} highlighting the agreement on the oscillatory features in the multipole moments for both the moment expansion and a cumulant expansion.  Figure \ref{fig:ind} shows the same excellent agreement for features induced by changes in the expansion history at high redshift, rather than imprinted upon the primordial power spectrum.

This extensive set of comparisons and tests imply that current perturbative models are up to the task of constraining models with features given suitably accurate redshift-space clustering data.  Future surveys, capable of operating over large volumes at high redshift --- shifting the non-linear scale to higher $k$ and the fundamental mode to lower $k$ --- would be ideal in providing such constraints.
We intend to return to the detectability of these features by different surveys \cite{Ballardini20}, and the impact of degeneracies, in a future paper.  However, it is clear that future surveys that probe large volumes at high redshift should be able to constrain both primordial features and the expansion history over a broad range of cosmic history (via its effect on growth).

\acknowledgments
We would like to thank Emanuele Castorina, Sergey Sibiryakov, and Marko Simonovi\'c for many useful discussions. We also thank  the  authors of ref.~\cite{Hill20} for making public their modification of the CLASS Boltzmann code including early dark energy.
S.C.\ is supported by the National Science Foundation Graduate Research Fellowship (Grant No.~DGE 1106400) and by the UC Berkeley Theoretical Astrophysics Center Astronomy and Astrophysics Graduate Fellowship.
M.W.\ is supported by the U.S. Department of Energy and the NSF.
This research has made use of NASA's Astrophysics Data System and the arXiv preprint server.
This research used resources of the National Energy Research Scientific Computing Center (NERSC), a U.S. Department of Energy Office of Science User Facility operated under Contract No. DE-AC02-05CH11231.

\appendix

\section{Saddle-Point Approximation for Nonlinear Dispersions}
\label{app:saddle}

In this appendix we expand upon the argument for selecting scale $q=\omega/k$ in the logarithmic case, and show it follows as a special case of a more general expression in the large $\omega$ limit.
Suppose we have
\begin{equation}
    \Delta P_X(k) = P_{\Lambda \rm CDM}(k) \sin( \omega \phi(k) )
\end{equation}
where $\omega$ is a constant tunable parameter such that $\omega \phi(k)$ is the (nonlinear) phase of the oscillating feature. We are interested in applying the saddle-point result for linear oscillations to this case. 

Around some wavenumber $k = k_0$ of interest, we can Taylor-expand the phase as
\begin{equation}
    \omega \phi(k) \simeq \omega\ \Big( \phi_0 + \phi_0'(k-k_0) + \frac{1}{2} \phi_0'' (k-k_0)^2 + ... \Big)
\end{equation}
where primes indicate derivatives with respect to $k$. Roughly, the linear approximation $\phi_L = \phi_0 + \phi_0' (k-k_0)$ will be good provided that $|k - k_0| < |\phi_0'/\phi_0''|$ independently of $\omega$. This also defines the window within which to a good approximation $d\phi/dk \approx \phi_0'$.

However, unless $\phi(k)$ is linear, this window is not guaranteed to cover all interesting $k$. To restrict our calculation to the interval around $k_0$ where the linear approximation is valid, let us multiply by a function $W(k,k_0)$ with support in the interval and which falls to zero away from $k_0$ with characteristic width $\sigma_k = |\phi_0'/\phi_0''|$. Then we can approximate $W(k,k_0) \sin(\omega \phi(k))$ as $W(k,k_0) \sin(\omega \phi_L(k))$ everywhere. In addition, for the saddle-point approximation to work we also require that the Fourier-transformed feature be sharply localized at some scale, which will be equal to  $ q_X =|\omega \phi_0'|$ in our case. This condition is controlled by $\omega$, since the number of cycles within our interval is $N_{\rm cyc} = (2\pi)^{-1} |\omega \phi_0'^2 / \phi_0''|$; for a given $\phi(k)$ the bigger $\omega$ is the better.

Following the above we have that the ``windowed'' feature $W(k,k_0)\ \Delta P_X(k)$ receives a damping factor well-approximated by $\Sigma^2(q_X = |\omega \phi_0'|).$ By the same logic we can fill out the oscillatory signal by sequence of windows $W(k,k_i)$ each with width $\sim |\phi_i'/\phi_i''|$ such that $\sum_i W(k,k_i) = 1$ at all $k$. Then, at any $k$ we have that the damping is given by $\Sigma^2(q_X =| \omega \phi_i' |)$, which will to a very good approximation be equal to $\phi'(k)$ from our previous arguments.

Let us apply this argument for the logarithmic features as an example. Setting $\phi(k) = \ln(k/k_\ast)$, we see that the width is set by $\sigma_k \sim k$ and that the requirement $N_{\rm cyc} \gg 1$ is equivalent to $\omega \gg 2\pi$. This limit is met by our fiducial $\omega = 10$ and by the frequencies explored in the BOSS data in ref.~\cite{Beutler19}, who also set a related criterion for the sharpness of primordial features. Note that the fact that the corresponding $\omega$'s are large also implies that the approximation $\phi'(k) \approx \phi_i'$ is a good one, with corrections of order $\omega^{-1}$ for fixed $N_{\rm cyc}$.

\section{GSM vs.\ moment expansion}
\label{app:gsm}

Beyond the moment expansion, the velocity statistics underlying redshift-space distortions can also be expanded via cumulant expansions to yield a variety of so-called streaming models \cite{VlaWhi19}. A popular example is the Gaussian streaming model (GSM) \cite{Reid11,VlaCasWhi16}, which derives from the cumulant expansion in (real) configuration space at second order. A particular strength of the GSM is its ability to accurately capture the nonlinear smoothing of the BAO feature in redshift space, which can be roughly attributed to the responsible bulk displacements truncating at second order in the configuration-space cumulants \cite{Chen20}.

Recently \cite{Chen20}, we argued that percent-level modeling of the redshift-space power spectrum requires including the third moment, or at least approximating its effect via a counterterm ansatz. Naively, this would rule out using the GSM for full-shape RSD analyses at intermediate (but perturbative) scales; however, a proper accounting of the counterterms in the second moment shows that its quadrupole requires a counterterm degenerate with the above ansatz. Specifically, splitting the two-point function into contributions from large-scale bias and effective corrections as in Eq.~5.1 of ref.~\cite{Chen20} and comparing with expressions for the velocity moments in LPT (Eqs.~4.11-15) one sees that the complete set of counterterms for the power spectrum can be obtained within the GSM by setting
\begin{equation}
    \alpha_P = \alpha_0, \quad \alpha_v = \alpha_2 - \frac{1}{18} \alpha_4,\quad \alpha^{(2)}_\sigma = - \frac{2}{9} \alpha_4,
\end{equation}
where $\alpha^{(2)}_\sigma$ is the counterterm to the quadrupole of $\sigma^2_{12}(\bk)$ multiplying the Zeldovich power spectrum $P_{\rm Zel}(k)$. This implies that the Fourier-transformed GSM can adequately model the redshift-space power spectrum while also accurately capturing nonlinear smoothing of power spectrum features.

\bibliographystyle{JHEP}
\bibliography{main}

\providecommand{\href}[2]{#2}\begingroup\raggedright\begin{thebibliography}{10}

\bibitem{Peacock99}
J.~A. {Peacock}, {\em {Cosmological Physics}}.
\newblock Jan., 1999.

\bibitem{Dodelson03}
S.~{Dodelson}, {\em {Modern cosmology}}.
\newblock 2003.

\bibitem{PlanckLegacy18}
{Planck Collaboration}, Y.~{Akrami}, F.~{Arroja}, M.~{Ashdown}, J.~{Aumont},
  C.~{Baccigalupi}, M.~{Ballardini}, A.~J. {Banday}, R.~B. {Barreiro},
  N.~{Bartolo}, S.~{Basak}, R.~{Battye}, K.~{Benabed}, J.~P. {Bernard},
  M.~{Bersanelli}, P.~{Bielewicz}, J.~J. {Bock}, J.~R. {Bond}, J.~{Borrill},
  F.~R. {Bouchet}, F.~{Boulanger}, M.~{Bucher}, C.~{Burigana}, R.~C. {Butler},
  E.~{Calabrese}, J.~F. {Cardoso}, J.~{Carron}, B.~{Casaponsa}, A.~{Challinor},
  H.~C. {Chiang}, L.~P.~L. {Colombo}, C.~{Combet}, D.~{Contreras}, B.~P.
  {Crill}, F.~{Cuttaia}, P.~{de Bernardis}, G.~{de Zotti}, J.~{Delabrouille},
  J.~M. {Delouis}, F.~X. {D{\'e}sert}, E.~{Di Valentino}, C.~{Dickinson}, J.~M.
  {Diego}, S.~{Donzelli}, O.~{Dor{\'e}}, M.~{Douspis}, A.~{Ducout}, X.~{Dupac},
  G.~{Efstathiou}, F.~{Elsner}, T.~A. {En{\ss}lin}, H.~K. {Eriksen},
  E.~{Falgarone}, Y.~{Fantaye}, J.~{Fergusson}, R.~{Fernandez-Cobos},
  F.~{Finelli}, F.~{Forastieri}, M.~{Frailis}, E.~{Franceschi}, A.~{Frolov},
  S.~{Galeotta}, S.~{Galli}, K.~{Ganga}, R.~T. {G{\'e}nova-Santos},
  M.~{Gerbino}, T.~{Ghosh}, J.~{Gonz{\'a}lez-Nuevo}, K.~M. {G{\'o}rski},
  S.~{Gratton}, A.~{Gruppuso}, J.~E. {Gudmundsson}, J.~{Hamann}, W.~{Hand ley},
  F.~K. {Hansen}, G.~{Helou}, D.~{Herranz}, E.~{Hivon}, Z.~{Huang}, A.~H.
  {Jaffe}, W.~C. {Jones}, A.~{Karakci}, E.~{Keih{\"a}nen}, R.~{Keskitalo},
  K.~{Kiiveri}, J.~{Kim}, T.~S. {Kisner}, L.~{Knox}, N.~{Krachmalnicoff},
  M.~{Kunz}, H.~{Kurki-Suonio}, G.~{Lagache}, J.~M. {Lamarre}, M.~{Langer},
  A.~{Lasenby}, M.~{Lattanzi}, C.~R. {Lawrence}, M.~{Le Jeune}, J.~P. {Leahy},
  J.~{Lesgourgues}, F.~{Levrier}, A.~{Lewis}, M.~{Liguori}, P.~B. {Lilje},
  M.~{Lilley}, V.~{Lindholm}, M.~{L{\'o}pez-Caniego}, P.~M. {Lubin}, Y.~Z.
  {Ma}, J.~F. {Mac{\'\i}as-P{\'e}rez}, G.~{Maggio}, D.~{Maino}, N.~{Mand
  olesi}, A.~{Mangilli}, A.~{Marcos-Caballero}, M.~{Maris}, P.~G. {Martin},
  E.~{Mart{\'\i}nez-Gonz{\'a}lez}, S.~{Matarrese}, N.~{Mauri}, J.~D. {McEwen},
  P.~D. {Meerburg}, P.~R. {Meinhold}, A.~{Melchiorri}, A.~{Mennella},
  M.~{Migliaccio}, M.~{Millea}, S.~{Mitra}, M.~A. {Miville-Desch{\^e}nes},
  D.~{Molinari}, A.~{Moneti}, L.~{Montier}, G.~{Morgante}, A.~{Moss},
  S.~{Mottet}, M.~{M{\"u}nchmeyer}, P.~{Natoli}, H.~U. {N{\o}rgaard-Nielsen},
  C.~A. {Oxborrow}, L.~{Pagano}, D.~{Paoletti}, B.~{Partridge}, G.~{Patanchon},
  T.~J. {Pearson}, M.~{Peel}, H.~V. {Peiris}, F.~{Perrotta}, V.~{Pettorino},
  F.~{Piacentini}, L.~{Polastri}, G.~{Polenta}, J.~L. {Puget}, J.~P. {Rachen},
  M.~{Reinecke}, M.~{Remazeilles}, A.~{Renzi}, G.~{Rocha}, C.~{Rosset},
  G.~{Roudier}, J.~A. {Rubi{\~n}o-Mart{\'\i}n}, B.~{Ruiz-Granados},
  L.~{Salvati}, M.~{Sandri}, M.~{Savelainen}, D.~{Scott}, E.~P.~S. {Shellard},
  M.~{Shiraishi}, C.~{Sirignano}, G.~{Sirri}, L.~D. {Spencer}, R.~{Sunyaev},
  A.~S. {Suur-Uski}, J.~A. {Tauber}, D.~{Tavagnacco}, M.~{Tenti}, L.~{Terenzi},
  L.~{Toffolatti}, M.~{Tomasi}, T.~{Trombetti}, J.~{Valiviita}, B.~{Van Tent},
  L.~{Vibert}, P.~{Vielva}, F.~{Villa}, N.~{Vittorio}, B.~D. {Wandelt}, I.~K.
  {Wehus}, M.~{White}, S.~D.~M. {White}, A.~{Zacchei}, and A.~{Zonca}, {\it
  {Planck 2018 results. I. Overview and the cosmological legacy of Planck}},
  {\em arXiv e-prints} (July, 2018) arXiv:1807.06205,
  [\href{http://arxiv.org/abs/1807.06205}{{\tt arXiv:1807.06205}}].

\bibitem{Chluba15}
J.~{Chluba}, J.~{Hamann}, and S.~P. {Patil}, {\it {Features and new physical
  scales in primordial observables: Theory and observation}},  {\em
  International Journal of Modern Physics D} {\bf 24} (June, 2015) 1530023,
  [\href{http://arxiv.org/abs/1505.01834}{{\tt arXiv:1505.01834}}].

\bibitem{Slosar19b}
A.~{Slosar}, X.~{Chen}, C.~{Dvorkin}, D.~{Meerburg}, B.~{Wallisch}, D.~{Green},
  and E.~{Silverstein}, {\it {Scratches from the Past: Inflationary Archaeology
  through Features in the Power Spectrum of Primordial Fluctuations}},  {\em
  \baas} {\bf 51} (May, 2019) 98, [\href{http://arxiv.org/abs/1903.09883}{{\tt
  arXiv:1903.09883}}].

\bibitem{Weinberg13}
D.~H. {Weinberg}, M.~J. {Mortonson}, D.~J. {Eisenstein}, C.~{Hirata}, A.~G.
  {Riess}, and E.~{Rozo}, {\it {Observational probes of cosmic acceleration}},
  {\em \physrep} {\bf 530} (Sept., 2013) 87--255,
  [\href{http://arxiv.org/abs/1201.2434}{{\tt arXiv:1201.2434}}].

\bibitem{PlanckInf18}
{Planck Collaboration}, Y.~{Akrami}, F.~{Arroja}, M.~{Ashdown}, J.~{Aumont},
  C.~{Baccigalupi}, M.~{Ballardini}, A.~J. {Banday}, R.~B. {Barreiro},
  N.~{Bartolo}, S.~{Basak}, K.~{Benabed}, J.~P. {Bernard}, M.~{Bersanelli},
  P.~{Bielewicz}, J.~J. {Bock}, J.~R. {Bond}, J.~{Borrill}, F.~R. {Bouchet},
  F.~{Boulanger}, M.~{Bucher}, C.~{Burigana}, R.~C. {Butler}, E.~{Calabrese},
  J.~F. {Cardoso}, J.~{Carron}, A.~{Challinor}, H.~C. {Chiang}, L.~P.~L.
  {Colombo}, C.~{Combet}, D.~{Contreras}, B.~P. {Crill}, F.~{Cuttaia}, P.~{de
  Bernardis}, G.~{de Zotti}, J.~{Delabrouille}, J.~M. {Delouis}, E.~{Di
  Valentino}, J.~M. {Diego}, S.~{Donzelli}, O.~{Dor{\'e}}, M.~{Douspis},
  A.~{Ducout}, X.~{Dupac}, S.~{Dusini}, G.~{Efstathiou}, F.~{Elsner}, T.~A.
  {En{\ss}lin}, H.~K. {Eriksen}, Y.~{Fantaye}, J.~{Fergusson},
  R.~{Fernandez-Cobos}, F.~{Finelli}, F.~{Forastieri}, M.~{Frailis},
  E.~{Franceschi}, A.~{Frolov}, S.~{Galeotta}, S.~{Galli}, K.~{Ganga},
  C.~{Gauthier}, R.~T. {G{\'e}nova-Santos}, M.~{Gerbino}, T.~{Ghosh},
  J.~{Gonz{\'a}lez-Nuevo}, K.~M. {G{\'o}rski}, S.~{Gratton}, A.~{Gruppuso},
  J.~E. {Gudmundsson}, J.~{Hamann}, W.~{Handley}, F.~K. {Hansen}, D.~{Herranz},
  E.~{Hivon}, D.~C. {Hooper}, Z.~{Huang}, A.~H. {Jaffe}, W.~C. {Jones},
  E.~{Keih{\"a}nen}, R.~{Keskitalo}, K.~{Kiiveri}, J.~{Kim}, T.~S. {Kisner},
  N.~{Krachmalnicoff}, M.~{Kunz}, H.~{Kurki-Suonio}, G.~{Lagache}, J.~M.
  {Lamarre}, A.~{Lasenby}, M.~{Lattanzi}, C.~R. {Lawrence}, M.~{Le Jeune},
  J.~{Lesgourgues}, F.~{Levrier}, A.~{Lewis}, M.~{Liguori}, P.~B. {Lilje},
  V.~{Lindholm}, M.~{Lpez-Caniego}, P.~M. {Lubin}, Y.~Z. {Ma}, J.~F.
  {Mac{\'\i}as-P{\'e}rez}, G.~{Maggio}, D.~{Maino}, N.~{Mandolesi},
  A.~{Mangilli}, A.~{Marcos-Caballero}, M.~{Maris}, P.~G. {Martin},
  E.~{Mart{\'\i}nez-Gonz{\'a}lez}, S.~{Matarrese}, N.~{Mauri}, J.~D. {McEwen},
  P.~D. {Meerburg}, P.~R. {Meinhold}, A.~{Melchiorri}, A.~{Mennella},
  M.~{Migliaccio}, S.~{Mitra}, M.~A. {Miville-Desch{\^e}nes}, D.~{Molinari},
  A.~{Moneti}, L.~{Montier}, G.~{Morgante}, A.~{Moss}, M.~{M{\"u}nchmeyer},
  P.~{Natoli}, H.~U. {N{\o}rgaard-Nielsen}, L.~{Pagano}, D.~{Paoletti},
  B.~{Partridge}, G.~{Patanchon}, H.~V. {Peiris}, F.~{Perrotta},
  V.~{Pettorino}, F.~{Piacentini}, L.~{Polastri}, G.~{Polenta}, J.~L. {Puget},
  J.~P. {Rachen}, M.~{Reinecke}, M.~{Remazeilles}, A.~{Renzi}, G.~{Rocha},
  C.~{Rosset}, G.~{Roudier}, J.~A. {Rubi{\~n}o-Mart{\'\i}n},
  B.~{Ruiz-Granados}, L.~{Salvati}, M.~{Sandri}, M.~{Savelainen}, D.~{Scott},
  E.~P.~S. {Shellard}, M.~{Shiraishi}, C.~{Sirignano}, G.~{Sirri}, L.~D.
  {Spencer}, R.~{Sunyaev}, A.~S. {Suur-Uski}, J.~A. {Tauber}, D.~{Tavagnacco},
  M.~{Tenti}, L.~{Toffolatti}, M.~{Tomasi}, T.~{Trombetti}, J.~{Valiviita},
  B.~{Van Tent}, P.~{Vielva}, F.~{Villa}, N.~{Vittorio}, B.~D. {Wandelt}, I.~K.
  {Wehus}, S.~D.~M. {White}, A.~{Zacchei}, J.~P. {Zibin}, and A.~{Zonca}, {\it
  {Planck 2018 results. X. Constraints on inflation}},  {\em arXiv e-prints}
  (July, 2018) arXiv:1807.06211, [\href{http://arxiv.org/abs/1807.06211}{{\tt
  arXiv:1807.06211}}].

\bibitem{Beutler19}
F.~{Beutler}, M.~{Biagetti}, D.~{Green}, A.~{Slosar}, and B.~{Wallisch}, {\it
  {Primordial features from linear to nonlinear scales}},  {\em Physical Review
  Research} {\bf 1} (Dec., 2019) 033209,
  [\href{http://arxiv.org/abs/1906.08758}{{\tt arXiv:1906.08758}}].

\bibitem{PCP18}
{Planck Collaboration}, N.~{Aghanim}, Y.~{Akrami}, M.~{Ashdown}, J.~{Aumont},
  C.~{Baccigalupi}, M.~{Ballardini}, A.~J. {Banday}, R.~B. {Barreiro},
  N.~{Bartolo}, S.~{Basak}, R.~{Battye}, K.~{Benabed}, J.~P. {Bernard},
  M.~{Bersanelli}, P.~{Bielewicz}, J.~J. {Bock}, J.~R. {Bond}, J.~{Borrill},
  F.~R. {Bouchet}, F.~{Boulanger}, M.~{Bucher}, C.~{Burigana}, R.~C. {Butler},
  E.~{Calabrese}, J.~F. {Cardoso}, J.~{Carron}, A.~{Challinor}, H.~C. {Chiang},
  J.~{Chluba}, L.~P.~L. {Colombo}, C.~{Combet}, D.~{Contreras}, B.~P. {Crill},
  F.~{Cuttaia}, P.~{de Bernardis}, G.~{de Zotti}, J.~{Delabrouille}, J.~M.
  {Delouis}, E.~{Di Valentino}, J.~M. {Diego}, O.~{Dor{\'e}}, M.~{Douspis},
  A.~{Ducout}, X.~{Dupac}, S.~{Dusini}, G.~{Efstathiou}, F.~{Elsner}, T.~A.
  {En{\ss}lin}, H.~K. {Eriksen}, Y.~{Fantaye}, M.~{Farhang}, J.~{Fergusson},
  R.~{Fernandez-Cobos}, F.~{Finelli}, F.~{Forastieri}, M.~{Frailis}, A.~A.
  {Fraisse}, E.~{Franceschi}, A.~{Frolov}, S.~{Galeotta}, S.~{Galli},
  K.~{Ganga}, R.~T. {G{\'e}nova-Santos}, M.~{Gerbino}, T.~{Ghosh},
  J.~{Gonz{\'a}lez-Nuevo}, K.~M. {G{\'o}rski}, S.~{Gratton}, A.~{Gruppuso},
  J.~E. {Gudmundsson}, J.~{Hamann}, W.~{Handley}, F.~K. {Hansen}, D.~{Herranz},
  S.~R. {Hildebrandt}, E.~{Hivon}, Z.~{Huang}, A.~H. {Jaffe}, W.~C. {Jones},
  A.~{Karakci}, E.~{Keih{\"a}nen}, R.~{Keskitalo}, K.~{Kiiveri}, J.~{Kim},
  T.~S. {Kisner}, L.~{Knox}, N.~{Krachmalnicoff}, M.~{Kunz}, H.~{Kurki-Suonio},
  G.~{Lagache}, J.~M. {Lamarre}, A.~{Lasenby}, M.~{Lattanzi}, C.~R. {Lawrence},
  M.~{Le Jeune}, P.~{Lemos}, J.~{Lesgourgues}, F.~{Levrier}, A.~{Lewis},
  M.~{Liguori}, P.~B. {Lilje}, M.~{Lilley}, V.~{Lindholm},
  M.~{L{\'o}pez-Caniego}, P.~M. {Lubin}, Y.~Z. {Ma}, J.~F.
  {Mac{\'\i}as-P{\'e}rez}, G.~{Maggio}, D.~{Maino}, N.~{Mandolesi},
  A.~{Mangilli}, A.~{Marcos-Caballero}, M.~{Maris}, P.~G. {Martin},
  M.~{Martinelli}, E.~{Mart{\'\i}nez-Gonz{\'a}lez}, S.~{Matarrese}, N.~{Mauri},
  J.~D. {McEwen}, P.~R. {Meinhold}, A.~{Melchiorri}, A.~{Mennella},
  M.~{Migliaccio}, M.~{Millea}, S.~{Mitra}, M.~A. {Miville-Desch{\^e}nes},
  D.~{Molinari}, L.~{Montier}, G.~{Morgante}, A.~{Moss}, P.~{Natoli}, H.~U.
  {N{\o}rgaard-Nielsen}, L.~{Pagano}, D.~{Paoletti}, B.~{Partridge},
  G.~{Patanchon}, H.~V. {Peiris}, F.~{Perrotta}, V.~{Pettorino},
  F.~{Piacentini}, L.~{Polastri}, G.~{Polenta}, J.~L. {Puget}, J.~P. {Rachen},
  M.~{Reinecke}, M.~{Remazeilles}, A.~{Renzi}, G.~{Rocha}, C.~{Rosset},
  G.~{Roudier}, J.~A. {Rubi{\~n}o-Mart{\'\i}n}, B.~{Ruiz-Granados},
  L.~{Salvati}, M.~{Sandri}, M.~{Savelainen}, D.~{Scott}, E.~P.~S. {Shellard},
  C.~{Sirignano}, G.~{Sirri}, L.~D. {Spencer}, R.~{Sunyaev}, A.~S. {Suur-Uski},
  J.~A. {Tauber}, D.~{Tavagnacco}, M.~{Tenti}, L.~{Toffolatti}, M.~{Tomasi},
  T.~{Trombetti}, L.~{Valenziano}, J.~{Valiviita}, B.~{Van Tent}, L.~{Vibert},
  P.~{Vielva}, F.~{Villa}, N.~{Vittorio}, B.~D. {Wand elt}, I.~K. {Wehus},
  M.~{White}, S.~D.~M. {White}, A.~{Zacchei}, and A.~{Zonca}, {\it {Planck 2018
  results. VI. Cosmological parameters}},  {\em arXiv e-prints} (July, 2018)
  arXiv:1807.06209, [\href{http://arxiv.org/abs/1807.06209}{{\tt
  arXiv:1807.06209}}].

\bibitem{Feng19}
Y.~{Feng}, M.-Y. {Chu}, U.~{Seljak}, and P.~{McDonald}, {\it {FastPM: Scaling
  N-body Particle Mesh solver}},  May, 2019.

\bibitem{nbdkit}
N.~{Hand}, Y.~{Feng}, F.~{Beutler}, Y.~{Li}, C.~{Modi}, U.~{Seljak}, and
  Z.~{Slepian}, {\it {nbodykit: An Open-source, Massively Parallel Toolkit for
  Large-scale Structure}},  {\em \aj} {\bf 156} (Oct, 2018) 160,
  [\href{http://arxiv.org/abs/1712.05834}{{\tt arXiv:1712.05834}}].

\bibitem{Grieb17}
J.~N. {Grieb}, A.~G. {S{\'a}nchez}, S.~{Salazar-Albornoz}, R.~{Scoccimarro},
  M.~{Crocce}, C.~{Dalla Vecchia}, F.~{Montesano}, H.~{Gil-Mar{\'\i}n}, A.~J.
  {Ross}, F.~{Beutler}, S.~{Rodr{\'\i}guez-Torres}, C.-H. {Chuang}, F.~{Prada},
  F.-S. {Kitaura}, A.~J. {Cuesta}, D.~J. {Eisenstein}, W.~J. {Percival},
  M.~{Vargas-Maga{\~n}a}, J.~L. {Tinker}, R.~{Tojeiro}, J.~R. {Brownstein},
  C.~{Maraston}, R.~C. {Nichol}, M.~D. {Olmstead}, L.~{Samushia}, H.-J. {Seo},
  A.~{Streblyanska}, and G.-b. {Zhao}, {\it {The clustering of galaxies in the
  completed SDSS-III Baryon Oscillation Spectroscopic Survey: Cosmological
  implications of the Fourier space wedges of the final sample}},  {\em \mnras}
  {\bf 467} (May, 2017) 2085--2112,
  [\href{http://arxiv.org/abs/1607.03143}{{\tt arXiv:1607.03143}}].

\bibitem{DESI}
{DESI Collaboration}, A.~{Aghamousa}, J.~{Aguilar}, S.~{Ahlen}, S.~{Alam},
  L.~E. {Allen}, C.~{Allende Prieto}, J.~{Annis}, S.~{Bailey}, C.~{Balland},
  and et~al., {\it {The DESI Experiment Part I: Science,Targeting, and Survey
  Design}},  {\em ArXiv e-prints} (Oct., 2016)
  [\href{http://arxiv.org/abs/1611.00036}{{\tt arXiv:1611.00036}}].

\bibitem{MegaMapper}
D.~{Schlegel}, J.~A. {Kollmeier}, and S.~{Ferraro}, {\it {The MegaMapper: a
  $z>2$ spectroscopic instrument for the study of Inflation and Dark Energy}},
  in {\em \baas}, vol.~51, p.~229, Sept., 2019.
\newblock \href{http://arxiv.org/abs/1907.11171}{{\tt arXiv:1907.11171}}.

\bibitem{MSE}
{The MSE Science Team}, C.~{Babusiaux}, M.~{Bergemann}, A.~{Burgasser},
  S.~{Ellison}, D.~{Haggard}, D.~{Huber}, M.~{Kaplinghat}, T.~{Li},
  J.~{Marshall}, S.~{Martell}, A.~{McConnachie}, W.~{Percival}, A.~{Robotham},
  Y.~{Shen}, S.~{Thirupathi}, K.-V. {Tran}, C.~{Yeche}, D.~{Yong},
  V.~{Adibekyan}, V.~{Silva Aguirre}, G.~{Angelou}, M.~{Asplund}, M.~{Balogh},
  P.~{Banerjee}, M.~{Bannister}, D.~{Barr{\'\i}a}, G.~{Battaglia}, A.~{Bayo},
  K.~{Bechtol}, P.~G. {Beck}, T.~C. {Beers}, E.~P. {Bellinger}, T.~{Berg},
  J.~M. {Bestenlehner}, M.~{Bilicki}, B.~{Bitsch}, J.~{Bland-Hawthorn}, A.~S.
  {Bolton}, A.~{Boselli}, J.~{Bovy}, A.~{Bragaglia}, D.~{Buzasi}, E.~{Caffau},
  J.~{Cami}, T.~{Carleton}, L.~{Casagrande}, S.~{Cassisi}, M.~{Catelan},
  C.~{Chang}, L.~{Cortese}, I.~{Damjanov}, L.~J.~M. {Davies}, R.~{de Grijs},
  G.~{de Rosa}, A.~{Deason}, P.~{di Matteo}, A.~{Drlica-Wagner}, D.~{Erkal},
  A.~{Escorza}, L.~{Ferrarese}, S.~W. {Fleming}, A.~{Font-Ribera},
  K.~{Freeman}, B.~T. {G{\"a}nsicke}, M.~{Gabdeev}, S.~{Gallagher},
  D.~{Gandolfi}, R.~A. {Garc{\'\i}a}, P.~{Gaulme}, M.~{Geha}, M.~{Gennaro},
  M.~{Gieles}, K.~{Gilbert}, Y.~{Gordon}, A.~{Goswami}, J.~P. {Greco},
  C.~{Grillmair}, G.~{Guiglion}, V.~{H{\'e}nault-Brunet}, P.~{Hall}, G.~{Hand
  ler}, T.~{Hansen}, N.~{Hathi}, D.~{Hatzidimitriou}, M.~{Haywood}, J.~V.
  {Hern{\'a}ndez Santisteban}, L.~{Hillenbrand}, A.~M. {Hopkins}, C.~{Howlett},
  M.~J. {Hudson}, R.~{Ibata}, D.~{Ili{\'c}}, P.~{Jablonka}, A.~{Ji},
  L.~{Jiang}, S.~{Juneau}, A.~{Karakas}, D.~{Karinkuzhi}, S.~Y. {Kim},
  X.~{Kong}, I.~{Konstantopoulos}, J.-K. {Krogager}, C.~{Lagos},
  R.~{Lallement}, C.~{Laporte}, Y.~{Lebreton}, K.-G. {Lee}, G.~F. {Lewis},
  S.~{Lianou}, X.~{Liu}, N.~{Lodieu}, J.~{Loveday}, S.~{M{\'e}sz{\'a}ros},
  M.~{Makler}, Y.-Y. {Mao}, D.~{Marchesini}, N.~{Martin}, M.~{Mateo},
  C.~{Melis}, T.~{Merle}, A.~{Miglio}, F.~{Gohar Mohammad},
  K.~{Molaverdikhani}, R.~{Monier}, T.~{Morel}, B.~{Mosser}, D.~{Nataf},
  L.~{Necib}, H.~R. {Neilson}, J.~A. {Newman}, A.~M. {Nierenberg}, B.~{Nord},
  P.~{Noterdaeme}, C.~{O'Dea}, M.~{Oshagh}, A.~B. {Pace},
  N.~{Palanque-Delabrouille}, G.~{Pandey}, L.~C. {Parker}, M.~S. {Pawlowski},
  A.~H.~G. {Peter}, P.~{Petitjean}, A.~{Petric}, V.~{Placco}, L.~{\v{C}}.
  {Popovi{\'c}}, A.~M. {Price-Whelan}, A.~{Prsa}, S.~{Ravindranath}, R.~M.
  {Rich}, J.~{Ruan}, J.~{Rybizki}, C.~{Sakari}, R.~E. {Sanderson},
  R.~{Schiavon}, C.~{Schimd}, A.~{Serenelli}, A.~{Siebert}, M.~{Siudek},
  R.~{Smiljanic}, D.~{Smith}, J.~{Sobeck}, E.~{Starkenburg}, D.~{Stello}, G.~M.
  {Szab{\'o}}, R.~{Szabo}, M.~A. {Taylor}, K.~{Thanjavur}, G.~{Thomas},
  E.~{Tollerud}, S.~{Toonen}, P.-E. {Tremblay}, L.~{Tresse}, M.~{Tsantaki},
  M.~{Valentini}, S.~{Van Eck}, A.~{Variu}, K.~{Venn}, E.~{Villaver}, M.~G.
  {Walker}, Y.~{Wang}, Y.~{Wang}, M.~J. {Wilson}, N.~{Wright}, S.~{Xu},
  M.~{Yildiz}, H.~{Zhang}, K.~{Zwintz}, B.~{Anguiano}, M.~{Bedell},
  W.~{Chaplin}, R.~{Collet}, J.-C. {Cuillandre}, P.-A. {Duc}, N.~{Flagey},
  J.~{Hermes}, A.~{Hill}, D.~{Kamath}, M.~B. {Laychak}, K.~{Ma{\l}ek},
  M.~{Marley}, A.~{Sheinis}, D.~{Simons}, S.~G. {Sousa}, K.~{Szeto}, Y.-S.
  {Ting}, S.~{Vegetti}, L.~{Wells}, F.~{Babas}, S.~{Bauman}, A.~{Bosselli},
  P.~{C{\^o}t{\'e}}, M.~{Colless}, J.~{Comparat}, H.~{Courtois}, D.~{Crampton},
  S.~{Croom}, L.~{Davies}, R.~{de Grijs}, K.~{Denny}, D.~{Devost}, P.~{di
  Matteo}, S.~{Driver}, M.~{Fernandez-Lorenzo}, R.~{Guhathakurta}, Z.~{Han},
  C.~{Higgs}, V.~{Hill}, K.~{Ho}, A.~{Hopkins}, M.~{Hudson}, R.~{Ibata},
  S.~{Isani}, M.~{Jarvis}, A.~{Johnson}, E.~{Jullo}, N.~{Kaiser}, J.-P.
  {Kneib}, J.~{Koda}, G.~{Koshy}, S.~{Mignot}, R.~{Murowinski}, J.~{Newman},
  A.~{Nusser}, A.~{Pancoast}, E.~{Peng}, C.~{Peroux}, C.~{Pichon},
  B.~{Poggianti}, J.~{Richard}, D.~{Salmon}, A.~{Seibert}, P.~{Shastri},
  D.~{Smith}, F.~{Sutaria}, C.~{Tao}, E.~{Taylor}, B.~{Tully}, L.~{van
  Waerbeke}, T.~{Vermeulen}, M.~{Walker}, J.~{Willis}, C.~{Willot}, and
  K.~{Withington}, {\it {The Detailed Science Case for the Maunakea
  Spectroscopic Explorer, 2019 edition}},  {\em arXiv e-prints} (Apr., 2019)
  arXiv:1904.04907, [\href{http://arxiv.org/abs/1904.04907}{{\tt
  arXiv:1904.04907}}].

\bibitem{Slosar19a}
A.~{Slosar}, Z.~{Ahmed}, D.~{Alonso}, M.~A. {Amin}, E.~J. {Arena},
  K.~{Bandura}, N.~{Battaglia}, J.~{Blazek}, P.~{Bull}, E.~{Castorina}, T.-C.
  {Chang}, L.~{Connor}, R.~{Dav{\'e}}, C.~{Dvorkin}, A.~{van Engelen},
  S.~{Ferraro}, R.~{Flauger}, S.~{Foreman}, J.~{Frisch}, D.~{Green},
  G.~{Holder}, D.~{Jacobs}, M.~C. {Johnson}, J.~S. {Dillon}, D.~{Karagiannis},
  A.~A. {Kaurov}, L.~{Knox}, A.~{Liu}, M.~{Loverde}, Y.-Z. {Ma}, K.~W. {Masui},
  T.~{McClintock}, K.~{Moodley}, M.~{Munchmeyer}, L.~B. {Newburgh}, C.~{Ng},
  A.~{Nomerotski}, P.~{O'Connor}, A.~{Obuljen}, H.~{Padmanabhan},
  D.~{Parkinson}, J.~X. {Prochaska}, S.~{Rajendran}, D.~{Rapetti},
  B.~{Saliwanchik}, E.~{Schaan}, N.~{Sehgal}, J.~R. {Shaw}, C.~{Sheehy},
  E.~{Sheldon}, R.~{Shirley}, E.~{Silverstein}, T.~{Slatyer}, A.~{Slosar},
  P.~{Stankus}, A.~{Stebbins}, P.~T. {Timbie}, G.~S. {Tucker}, W.~{Tyndall},
  F.~{Villaescusa Navarro}, B.~{Wallisch}, and M.~{White}, {\it {Packed
  Ultra-wideband Mapping Array (PUMA): A Radio Telescope for Cosmology and
  Transients}},  in {\em \baas}, vol.~51, p.~53, Sep, 2019.
\newblock \href{http://arxiv.org/abs/1907.12559}{{\tt arXiv:1907.12559}}.

\bibitem{Adams01}
J.~{Adams}, B.~{Cresswell}, and R.~{Easther}, {\it {Inflationary perturbations
  from a potential with a step}},  {\em \prd} {\bf 64} (Dec., 2001) 123514,
  [\href{http://xxx.lanl.gov/abs/astro-ph/0102236}{{\tt astro-ph/0102236}}].

\bibitem{Hill20}
J.~C. {Hill}, E.~{McDonough}, M.~W. {Toomey}, and S.~{Alexander}, {\it {Early
  Dark Energy Does Not Restore Cosmological Concordance}},  {\em arXiv
  e-prints} (Mar., 2020) arXiv:2003.07355,
  [\href{http://arxiv.org/abs/2003.07355}{{\tt arXiv:2003.07355}}].

\bibitem{Ivanov20}
M.~M. {Ivanov}, E.~{McDonough}, J.~C. {Hill}, M.~{Simonovi{\'c}}, M.~W.
  {Toomey}, S.~{Alexand er}, and M.~{Zaldarriaga}, {\it {Constraining Early
  Dark Energy with Large-Scale Structure}},  {\em arXiv e-prints} (June, 2020)
  arXiv:2006.11235, [\href{http://arxiv.org/abs/2006.11235}{{\tt
  arXiv:2006.11235}}].

\bibitem{DAmico20b}
G.~{D'Amico}, L.~{Senatore}, P.~{Zhang}, and H.~{Zheng}, {\it {The Hubble
  Tension in Light of the Full-Shape Analysis of Large-Scale Structure Data}},
  {\em arXiv e-prints} (June, 2020) arXiv:2006.12420,
  [\href{http://arxiv.org/abs/2006.12420}{{\tt arXiv:2006.12420}}].

\bibitem{Klypin20}
A.~{Klypin}, V.~{Poulin}, F.~{Prada}, J.~{Primack}, M.~{Kamionkowski},
  V.~{Avila-Reese}, A.~{Rodriguez-Puebla}, P.~{Behroozi}, D.~{Hellinger}, and
  T.~L. {Smith}, {\it {Clustering and Halo Abundances in Early Dark Energy
  Cosmological Models}},  {\em arXiv e-prints} (June, 2020) arXiv:2006.14910,
  [\href{http://arxiv.org/abs/2006.14910}{{\tt arXiv:2006.14910}}].

\bibitem{Smith19}
V.~{Poulin}, T.~L. {Smith}, T.~{Karwal}, and M.~{Kamionkowski}, {\it {Early
  Dark Energy can Resolve the Hubble Tension}},  {\em \prl} {\bf 122} (June,
  2019) 221301, [\href{http://arxiv.org/abs/1811.04083}{{\tt
  arXiv:1811.04083}}].

\bibitem{Smith20}
T.~L. {Smith}, V.~{Poulin}, and M.~A. {Amin}, {\it {Oscillating scalar fields
  and the Hubble tension: A resolution with novel signatures}},  {\em \prd}
  {\bf 101} (Mar., 2020) 063523, [\href{http://arxiv.org/abs/1908.06995}{{\tt
  arXiv:1908.06995}}].

\bibitem{CLASS}
D.~{Blas}, J.~{Lesgourgues}, and T.~{Tram}, {\it {The Cosmic Linear Anisotropy
  Solving System (CLASS). Part II: Approximation schemes}},  {\em \jcap} {\bf
  7} (July, 2011) 034, [\href{http://arxiv.org/abs/1104.2933}{{\tt
  arXiv:1104.2933}}].

\bibitem{Bha96}
S.~{Bharadwaj}, {\it {The Evolution of Correlation Functions in the Zeldovich
  Approximation and Its Implications for the Validity of Perturbation Theory}},
   {\em \apj} {\bf 472} (Nov., 1996) 1--+,
  [\href{http://arxiv.org/abs/astro-ph/9}{{\tt astro-ph/9}}].

\bibitem{Mat08a}
T.~{Matsubara}, {\it {Resumming cosmological perturbations via the Lagrangian
  picture: One-loop results in real space and in redshift space}},  {\em \prd}
  {\bf 77} (Mar., 2008) 063530, [\href{http://arxiv.org/abs/0711.2521}{{\tt
  arXiv:0711.2521}}].

\bibitem{Mat08b}
T.~{Matsubara}, {\it {Nonlinear perturbation theory with halo bias and
  redshift-space distortions via the Lagrangian picture}},  {\em \prd} {\bf 78}
  (Oct., 2008) 083519, [\href{http://arxiv.org/abs/0807.1733}{{\tt
  arXiv:0807.1733}}].

\bibitem{ESW07}
D.~J. {Eisenstein}, H.-J. {Seo}, and M.~{White}, {\it {On the Robustness of the
  Acoustic Scale in the Low-Redshift Clustering of Matter}},  {\em \apj} {\bf
  664} (Aug., 2007) 660--674,
  [\href{http://xxx.lanl.gov/abs/astro-ph/0604361}{{\tt astro-ph/0604361}}].

\bibitem{PWC09}
N.~{Padmanabhan}, M.~{White}, and J.~D. {Cohn}, {\it {Reconstructing baryon
  oscillations: A Lagrangian theory perspective}},  {\em \prd} {\bf 79} (Mar.,
  2009) 063523, [\href{http://arxiv.org/abs/0812.2905}{{\tt arXiv:0812.2905}}].

\bibitem{Noh09}
Y.~{Noh}, M.~{White}, and N.~{Padmanabhan}, {\it {Reconstructing baryon
  oscillations}},  {\em \prd} {\bf 80} (Dec., 2009) 123501,
  [\href{http://arxiv.org/abs/0909.1802}{{\tt arXiv:0909.1802}}].

\bibitem{CLPT}
J.~{Carlson}, B.~{Reid}, and M.~{White}, {\it {Convolution Lagrangian
  perturbation theory for biased tracers}},  {\em \mnras} {\bf 429} (Feb.,
  2013) 1674--1685, [\href{http://arxiv.org/abs/1209.0780}{{\tt
  arXiv:1209.0780}}].

\bibitem{TasZal12}
S.~{Tassev} and M.~{Zaldarriaga}, {\it {Towards an optimal reconstruction of
  baryon oscillations}},  {\em \jcap} {\bf 10} (Oct., 2012) 006,
  [\href{http://arxiv.org/abs/1203.6066}{{\tt arXiv:1203.6066}}].

\bibitem{McCSza12}
N.~{McCullagh} and A.~S. {Szalay}, {\it {Nonlinear Behavior of Baryon Acoustic
  Oscillations from the Zel'dovich Approximation Using a Non-Fourier
  Perturbation Approach}},  {\em \apj} {\bf 752} (June, 2012) 21,
  [\href{http://arxiv.org/abs/1202.1306}{{\tt arXiv:1202.1306}}].

\bibitem{White14}
M.~{White}, {\it {The Zel'dovich approximation}},  {\em \mnras} {\bf 439}
  (Apr., 2014) 3630--3640, [\href{http://arxiv.org/abs/1401.5466}{{\tt
  arXiv:1401.5466}}].

\bibitem{Vlah16}
Z.~{Vlah}, U.~{Seljak}, M.~{Yat Chu}, and Y.~{Feng}, {\it {Perturbation theory,
  effective field theory, and oscillations in the power spectrum}},  {\em
  \jcap} {\bf 2016} (Mar, 2016) 057,
  [\href{http://arxiv.org/abs/1509.02120}{{\tt arXiv:1509.02120}}].

\bibitem{McQuinn16}
M.~{McQuinn} and M.~{White}, {\it {Cosmological perturbation theory in 1+1
  dimensions}},  {\em \jcap} {\bf 1} (Jan., 2016) 043,
  [\href{http://arxiv.org/abs/1502.07389}{{\tt arXiv:1502.07389}}].

\bibitem{Crocce08}
M.~{Crocce} and R.~{Scoccimarro}, {\it {Nonlinear evolution of baryon acoustic
  oscillations}},  {\em \prd} {\bf 77} (Jan., 2008) 023533,
  [\href{http://arxiv.org/abs/0704.2783}{{\tt arXiv:0704.2783}}].

\bibitem{SenZal15}
L.~{Senatore} and M.~{Zaldarriaga}, {\it {The IR-resummed Effective Field
  Theory of Large Scale Structures}},  {\em \jcap} {\bf 2} (Feb., 2015) 13,
  [\href{http://arxiv.org/abs/1404.5954}{{\tt arXiv:1404.5954}}].

\bibitem{Schmittfull15}
M.~{Schmittfull}, Y.~{Feng}, F.~{Beutler}, B.~{Sherwin}, and M.~Y. {Chu}, {\it
  {Eulerian BAO reconstructions and N -point statistics}},  {\em \prd} {\bf 92}
  (Dec., 2015) 123522, [\href{http://arxiv.org/abs/1508.06972}{{\tt
  arXiv:1508.06972}}].

\bibitem{Baldauf15}
T.~{Baldauf}, M.~{Mirbabayi}, M.~{Simonovi{\'c}}, and M.~{Zaldarriaga}, {\it
  {Equivalence principle and the baryon acoustic peak}},  {\em \prd} {\bf 92}
  (Aug., 2015) 043514, [\href{http://arxiv.org/abs/1504.04366}{{\tt
  arXiv:1504.04366}}].

\bibitem{Blas16}
D.~{Blas}, M.~{Garny}, M.~M. {Ivanov}, and S.~{Sibiryakov}, {\it {Time-sliced
  perturbation theory II: baryon acoustic oscillations and infrared
  resummation}},  {\em \jcap} {\bf 2016} (July, 2016) 028,
  [\href{http://arxiv.org/abs/1605.02149}{{\tt arXiv:1605.02149}}].

\bibitem{Seo16}
H.-J. {Seo}, F.~{Beutler}, A.~J. {Ross}, and S.~{Saito}, {\it {Modeling the
  reconstructed BAO in Fourier space}},  {\em \mnras} {\bf 460} (Aug., 2016)
  2453--2471, [\href{http://arxiv.org/abs/1511.00663}{{\tt arXiv:1511.00663}}].

\bibitem{Ding17}
Z.~{Ding}, H.-J. {Seo}, Z.~{Vlah}, Y.~{Feng}, M.~{Schmittfull}, and
  F.~{Beutler}, {\it {Theoretical systematics of Future Baryon Acoustic
  Oscillation Surveys}},  {\em \mnras} {\bf 479} (Sept., 2018) 1021--1054,
  [\href{http://arxiv.org/abs/1708.01297}{{\tt arXiv:1708.01297}}].

\bibitem{Hikage17}
C.~{Hikage}, K.~{Koyama}, and A.~{Heavens}, {\it {Perturbation theory for BAO
  reconstructed fields: One-loop results in the real-space matter density
  field}},  {\em \prd} {\bf 96} (Aug., 2017) 043513,
  [\href{http://arxiv.org/abs/1703.07878}{{\tt arXiv:1703.07878}}].

\bibitem{Peloso17}
M.~{Peloso} and M.~{Pietroni}, {\it {Galilean invariant resummation schemes of
  cosmological perturbations}},  {\em \jcap} {\bf 2017} (Jan., 2017) 056,
  [\href{http://arxiv.org/abs/1609.06624}{{\tt arXiv:1609.06624}}].

\bibitem{Ivanov18}
M.~M. {Ivanov} and S.~{Sibiryakov}, {\it {Infrared resummation for biased
  tracers in redshift space}},  {\em \jcap} {\bf 2018} (July, 2018) 053,
  [\href{http://arxiv.org/abs/1804.05080}{{\tt arXiv:1804.05080}}].

\bibitem{Vasudevan19}
A.~{Vasudevan}, M.~M. {Ivanov}, S.~{Sibiryakov}, and J.~{Lesgourgues}, {\it
  {Time-sliced perturbation theory with primordial non-Gaussianity and effects
  of large bulk flows on inflationary oscillating features}},  {\em \jcap} {\bf
  2019} (Sept., 2019) 037, [\href{http://arxiv.org/abs/1906.08697}{{\tt
  arXiv:1906.08697}}].

\bibitem{Ballardini20}
M.~{Ballardini}, R.~{Murgia}, M.~{Baldi}, F.~{Finelli}, and M.~{Viel}, {\it
  {Non-linear damping of superimposed primordial oscillations on the matter
  power spectrum in galaxy surveys}},  {\em \jcap} {\bf 2020} (Apr., 2020) 030,
  [\href{http://arxiv.org/abs/1912.12499}{{\tt arXiv:1912.12499}}].

\bibitem{Chen20}
S.-F. {Chen}, Z.~{Vlah}, and M.~{White}, {\it {Consistent Modeling of Velocity
  Statistics and Redshift-Space Distortions in One-Loop Perturbation Theory}},
  {\em arXiv e-prints} (May, 2020) arXiv:2005.00523,
  [\href{http://arxiv.org/abs/2005.00523}{{\tt arXiv:2005.00523}}].

\bibitem{Lazeyras16}
T.~{Lazeyras}, M.~{Musso}, and V.~{Desjacques}, {\it {Lagrangian bias of
  generic large-scale structure tracers}},  {\em \prd} {\bf 93} (Mar., 2016)
  063007, [\href{http://arxiv.org/abs/1512.05283}{{\tt arXiv:1512.05283}}].

\bibitem{Abidi18}
M.~M. {Abidi} and T.~{Baldauf}, {\it {Cubic halo bias in Eulerian and
  Lagrangian space}},  {\em \jcap} {\bf 2018} (July, 2018) 029,
  [\href{http://arxiv.org/abs/1802.07622}{{\tt arXiv:1802.07622}}].

\bibitem{VlaWhi19}
Z.~{Vlah} and M.~{White}, {\it {Exploring redshift-space distortions in
  large-scale structure}},  {\em \jcap} {\bf 2019} (Mar, 2019) 007,
  [\href{http://arxiv.org/abs/1812.02775}{{\tt arXiv:1812.02775}}].

\bibitem{Reid11}
B.~A. {Reid} and M.~{White}, {\it {Towards an accurate model of the
  redshift-space clustering of haloes in the quasi-linear regime}},  {\em
  \mnras} {\bf 417} (Nov., 2011) 1913--1927,
  [\href{http://arxiv.org/abs/1105.4165}{{\tt arXiv:1105.4165}}].

\bibitem{VlaCasWhi16}
Z.~{Vlah}, E.~{Castorina}, and M.~{White}, {\it {The Gaussian streaming model
  and convolution Lagrangian effective field theory}},  {\em \jcap} {\bf 12}
  (Dec., 2016) 007, [\href{http://arxiv.org/abs/1609.02908}{{\tt
  arXiv:1609.02908}}].

\end{thebibliography}\endgroup

\end{document}